\documentclass[aps,twocolumn,prd,superscriptaddress,showpacs,preprintnumbers,amsmath,amssymb,nofootinbib]{revtex4-1}
\pdfoutput=1

\usepackage{amsmath}
\usepackage{amsfonts}
\usepackage{amssymb}
\usepackage{bbm}
\usepackage{dsfont}

\usepackage{graphicx}
\usepackage{subfigure}
\usepackage{slashed}% for slashed
\usepackage{color}
\usepackage{xspace}
\usepackage{hyperref}

\newcommand{\imag}{\text{i}}
\newcommand{\skipthis}[1]{}
\renewcommand{\d}{{\text{d}}}
\newcommand{\Tr}{{\text{Tr}}}
\newcommand{\qcd}{QC\ensuremath{_2}D\xspace}
\newcommand{\Ur}{\ensuremath{U_{k,\rho}}}
\newcommand{\Urr}{\ensuremath{U_{k,\rho\rho}}}
\newcommand{\Ud}{\ensuremath{U_{k,d}}}
\newcommand{\Up}{\ensuremath{U_{k,\phi}}}
\newcommand{\Upp}{\ensuremath{U_{k,\phi\phi}}}
\newcommand{\Udd}{\ensuremath{U_{k,dd}}}
\newcommand{\Urd}{\ensuremath{U_{k,\rho d}}} 
 \def\Eq#1{Eq.~(\ref{#1})}
 
\def\Eqs#1{Eqs.~(\ref{#1})}

\newcommand{\smallfrac}[2]{\mbox{\small ${\displaystyle \frac{#1}{#2}}$}}
\renewcommand{\log}{\ln}

\begin{document}

\title{Quark-meson-diquark model for two-color QCD}
% \title{\texttt{
% $-\,$Id: twocolorqm.tex 132 2012-02-28 07:39:32Z smekal $-$}\\
% Quark-meson-diquark model for two-color QCD}

\author{Nils Strodthoff}
\email{nstrodt@theorie.ikp.physik.tu-darmstadt.de} 
\affiliation{Institut f\"{u}r Kernphysik, Technische Universit\"{a}t Darmstadt, D-64289 Darmstadt, Germany}
\author{Bernd-Jochen Schaefer}
\email{bernd-jochen.schaefer@uni-graz.at}
\affiliation{Institut f\"{u}r Physik, Karl-Franzens-Universit\"{a}t Graz, A-8010 Graz, Austria}  
\affiliation{Institut f\"{u}r Theoretische Physik,
  Justus-Liebig-Universit\"{a}t Gie\ss en, D-35392 Gie\ss en, Germany}  
\author{Lorenz von Smekal}
\email{lorenz.smekal@physik.tu-darmstadt.de}
 \affiliation{Institut f\"{u}r Kernphysik, Technische Universit\"{a}t Darmstadt, D-64289 Darmstadt, Germany}
		%Lattice QCD calculations (see also 11.15.Ha
                        %Lattice gauge theory)
\begin{abstract}
  We introduce a two-flavor quark-meson-diquark model for two-color
  QCD and its extensions to include gauge-field dynamics as described
  by the Polyakov loop. Grand potential and phase structure are being
  studied both in mean-field approximation and with the functional
  renormalization group. The model provides an explicit example for
  the importance of baryonic degrees of freedom: When they are
  omitted, the phase diagram closely resembles that of the
  corresponding (Polyakov)-quark-meson models for QCD, in particular
  including their critical endpoint. In order to reproduce the well
  established main features based on the symmetries and breaking
  patterns of two-color QCD, however, they must be included and there
  is no critical endpoint. The competing dynamics of collective
  mesonic and baryonic fluctuations is well described by the
  functional renormalization group equation in lowest order derivative
  expansion for the effective potential which we solve numerically on
  a two-dimensional grid in field space.
\end{abstract}
\maketitle

%%%%%%%%%%%%%%%%%%%%%%%%%%%%%%%%%%%%%%%%%%%%%%%%%%%%%%%%%%%%%%%%%%%%%%%%%%%%%
\section{Introduction}

The phase diagram of Quantum Chromodynamics (QCD) is subject to
enormous international research campaigns \cite{BraunMunzinger:2009zz,
  Friman:2011zz}. In order to understand its main characteristic
features such as the different phases of strongly interacting matter,
the nature of the transitions between them, the existence and
locations of critical points or perhaps approximate triple points and
even multicritical points, it has proven to be very useful to deform
QCD by not only varying the individual quarks' masses but also the
numbers of their different flavors and colors.  An important example
is the limit of infinitely many colors $N_c$ which inspired many
qualitative descriptions of the QCD phase diagram
\cite{McLerran:2007qj,Hidaka:2008yy,Andronic:2009gj}.  One interesting
aspect of this limit is that the baryon density becomes an order
parameter for $N_c\to\infty$, in particular, also when the number of
flavors $N_f$ grows along with $N_c$, {\it i.e.} for $N_f/N_c$ held
fixed. In this paper we study two-color QCD, which shares this aspect
of the large-$N$ limit, here with $N_c=N_f=2$. If one accepts that the
phases of many-color QCD with  $N_c\sim N_f \to\infty$ have a bearing
on the real world, it might therefore not be absurd, with due
appreciation of all differences, to consider $N_c=N_f=2$ either.

Quantum Chromodynamics with two colors (\qcd) has been well studied
for many years within chiral effective field theory and random matrix theory
\cite{Kogut:1999iv, Kogut:2000ek, Splittorff:2000mm,
  Splittorff:2001fy, Dunne:2002vb, Brauner:2006dv, Kanazawa:2009ks,
  Kanazawa:2009en,Kanazawa:2011tt}, in lattice simulations
\cite{Nakamura:1984uz,Hands:1999md, Hands:2000ei, Muroya:2002jj,
  Chandrasekharan:2006tz, Hands:2006ve, Hands:2010gd, Hands:2011ye}, and the
Nambu--Jona-Lasinio model
\cite{Kondratyuk:1991hf, Kondratyuk:1992he, Rapp:1997zu, Ratti:2004ra,
  Sun:2007fc, Brauner:2009gu, Andersen:2010vu, Harada:2010vy, Zhang:2010kn, 
  He:2010nb}. In this paper we formulate a
Polyakov-quark-meson-diquark (PQMD) Model for
studying the phase diagram of QC$_2$D with the functional
renormalization group, including fluctuations due to collective
excitations.

The most important differences between two and three colors follow
from the special property of the $SU(2)$ gauge group of \qcd:
Its representations are either pseudo-real or real which leads to an
anti\-unitary symmetry in the Dirac operator \cite{Kogut:2000ek}. As a
result, the fermion determinant remains real for non-vanishing baryon
or quark chemical potential, $\mu\not=0$, as it does for adjoint
quarks in any-color QCD, or in the $G_2$ gauge theory with fundamental
fermions also, for example. Thus, for an even number of degenerate
fundamental quark flavors in \qcd there is {\em no fermion-sign problem} and
the phase diagram is amenable to lattice Monte-Carlo
simulations. Symmetry considerations, lattice simulations and
non-perturbative functional continuum methods together should
therefore allow us to understand the phase diagram of this theory
completely. A combined effort towards this goal will be very
worthwhile in particular because it will help to bring the functional
continuum methods to a level at which they can reliably be applied,
with the necessary adjustments, also to real QCD where lattice
simulations suffer from the infamous fermion-sign problem
\cite{deForcrand:2010ys}.

Another consequence of the pseudo-reality is the
Pauli-G\"ursey symmetry which allows to combine quarks and
charge-conjugated antiquarks into enlarged flavor multiplets. As a
result, for vanishing chemical potential and quark mass, $\mu=m_q=0$,
the usual $SU(N_f)\times SU(N_f) \times U(1)_B$ chiral and baryon
number symmetries are replaced by an extended $SU(2N_f)$ flavor
symmetry which is (spontaneously) broken by a (dynamical) Dirac mass
down to the $(2N_f+1)N_f$ dimensional compact symplectic
group $Sp(N_f)$.  For $N_f=2$ the extended flavor
symmetry group $SU(4)$ and its $Sp(2)$ subgroup are locally isomorphic
to the rotation groups $SO(6)$ and $SO(5)$, respectively. The coset is
given by $S^5$, the unit sphere in six dimensions, and a spontaneously generated
Dirac mass will lead to five Goldstone bosons, the three pions plus a
scalar diquark-antidiquark pair.

Moreover, for $N_c=2$ these color-singlet scalar diquarks play a dual
role as bosonic baryons at the same time. While this thus represents
the perhaps most important  
difference as compared to the real world, it also makes it much easier
to investigate the effects of baryonic degrees of freedom on the phase
diagram in functional approaches. In that sense our model can be
considered as a first step towards their inclusion in a
`quark-meson-baryon' model for real QCD with three colors.

For the same reason our model of \qcd provides a relativistic
analogue of the BEC-BCS crossover in ultracold fermionic
quantum-gases, which has also been described successfully with
functional renormalization group methods
\cite{Diehl:2009ma,Scherer:2010sv}.  
In contrast to non-relativistic models of the BEC-BCS crossover, 
an interesting additional constraint thereby arises from the Silver
Blaze property \cite{Cohen:2003kd}: When a relativistic chemical
potential $\mu$ is coupled to degrees of freedom with a mass gap
$\Delta$, at zero temperature, the partition function and hence
thermodynamic observables must actually remain independent of the
chemical potential as long as $\mu < \Delta$. We will see that it is
not trivial, in general, to implement this constraint in
non-perturbative functional renormalization group studies, and that it
can provide valuable extra information to devise intelligent truncations.

Our main interest here, however, is to explicitly demonstrate the
impact of baryonic degrees of freedom on the phase diagram by
comparing the purely mesonic model, representative of typical
three-color QCD model calculations, to the full quark-meson-diquark
model.

For this comparison we argue that it is more appropriate to think of
the vacuum diquark mass as the baryon mass $m_B$ rather than the pion
mass $m_\pi$.  In \qcd with its extended flavor symmetry they are
the same, but the essential aspect of this assignment is that a
continuous phase transition at zero temperature occurs at a critical
quark chemical potential $\mu_c = m_B/N_c$. Except for the scale
separation between $m_\pi$ and $m_B$ in the real world, this
transition can then be thought to correspond to the liquid-gas
transition of nuclear matter in QCD with three colors which is of
first order, involves the binding energy, and thus occurs somewhat
below $\mu = m_B/N_c$.

As temperature increases the liquid gas transition ends, turning into
a crossover with continuously varying but nevertheless probably still
relatively abruptly increasing baryon density along some narrow
region. This rapid increase is generally expected to lead to the
strong chemical-potential dependence of the chemical freeze-out line
observed in heavy ion collisions at center-of-mass energies below
about 10 GeV per nucleon pair, the baryonic freeze-out
\cite{BraunMunzinger:2007zz,Andronic:2008gu}. 
One might conclude that the phase transition line for diquark 
condensation, where a rapidly increasing baryon density spontaneously 
develops, would be the origin of a corresponding baryonic freeze-out line 
in two-color QCD, with $N_c=N_f =2$ arguably not
  necessarily further from reality than the large $N_c $ limits.  
As in the latter, one might then even identify a two-color version of
quarkyonic matter \cite{McLerran:2007qj, Hidaka:2008yy, Hands:2010gd,
  Brauner:2009gu}.

Finally, we would like to point out that our model, the functional
renormalization group equations and the techniques to solve them have
a broad scope of applications beyond two-color QCD. One example is QCD
with two light flavors at finite isospin chemical potential, which has
been studied with the NJL model in mean-field plus
random phase approximation (RPA) \cite{He:2005nk, Xiong:2009zz}. There
is a precise equivalence between the corresponding quark-meson model
with isospin chemical potential and 
our quark-meson-diquark model of two-color QCD. Besides changing $N_c$
this merely involves reducing the number of would-be Goldstone bosons
from five to three again, retaining only one of our degenerate pions,
and reinterpreting the diquarks as the charged pions with isospin
chemical potential \cite{KazuhikoEtAlInPrep}.  Similar models are also
studied in the context of color superconductivity
\cite{Alford:2007xm,Steiner:2002gx,Huang:2003xd}. The capacity to
numerically solve functional renormalization group equations 
on higher dimensional grids in field space is generally useful for
competing symmetries, as in a quark-meson model study of the axial
anomaly with scale dependent 't~Hooft couplings, for example
\cite{Mario2011}.  

The outline of this paper is as follows: In Sec.~\ref{sec:ConstrQM} we
review the general features of \qcd such as its enlarged flavor
symmetry and the possible symmetry breaking patterns in some more
detail. Based on these symmetry considerations we then construct the
Polyakov-loop extended quark-meson-diquark Lagrangian for \qcd. In the
next section, Sec.~\ref{sec:MF}, we derive the thermodynamic potential
of the (P)QMD model in mean-field approximation, discuss so-called
vacuum contributions, the Silver Blaze property and the relevance of
pole versus screening masses for mesons and diquarks.  The functional
renormalization group flow equations for the effective potential in
leading-order derivative expansion are derived in
Sec.~\ref{sec:funRG}. In this section 
we also calculate critical exponents which are consistent with the expected
symmetry breaking pattern, investigate in how far mean-field results
are modified by fluctuations, and give a transparent illustration of
the importance of baryonic degrees of freedom for the phase
diagram. As a byproduct we note that starting from a tricritical
point, a region of first-order transition limiting the diquark
condensation phase at larger chemical potentials as predicted from
chiral perturbation theory at next-to-leading order
\cite{Splittorff:2001fy}, is also observed in the QMD model at the
mean-field level. This first-order transition turns out to be a
mean-field artifact, however. It is washed out by the fluctuations,
and there is no sign of a tricritical point left, once the
thermodynamic potential is obtained from its functional
renormalization group flow. We draw our conclusions and present an
outlook in Sec.~\ref{sec:outlook}. Technical details can be found in
several appendices.

%%%%%%%%%%%%%%%%%%%%%%%%%%%%%%%%%%%%%%%%%%%%%%%%%%%%%%%%%%%%%%%%%%%%%%%%%%%%%
% Sec II intro PQMD model
%%%%%%%%%%%%%%%%%%%%%%%%%%%%%%%%%%%%%%%%%%%%%%%%%%%%%%%%%%%%%%%%%%%%%%%%%%%%%
\section{Flavor symmetries in \qcd  \label{sec:ConstrQM}}

We begin this section with a short review of the extended flavor
symmetries of \qcd due to its Pauli-G\"ursey symmetry, and the
associated symmetry breaking patterns. We then discuss a qualitative
phase diagram for two-flavor QC$_2$D and construct the
quark-meson-diquark (QMD) model by a suitable vector coupling of
quark bilinears to meson and diquark fields. 

%%%%%%%%%%%%%%%%%%%%%%%%%%%%%%%%%%%%%%%%%%%%%%%%%%%%%%%%%%%%%%%%%%%%%%%%%%%%%
%\subsection{Pauli-G\"ursey symmetry and  symmetry
% breaking patterns\label{sec:chsymbreaking}}
\subsection{Extended flavor  symmetries and symmetry 
 breaking patterns\label{sec:chsymbreaking}}

As all half-odd integer representations of $SU(2)$, its fundamental
representation is pseudoreal, which means that it is isomorphic to its
complex conjugate representation with the isometry given by $S =
\imag\sigma_2$, $S^2 = -1$.\footnote{The irreducible representations of
  the proper rotations are real which means their complex conjugates
  are obtained from isometries $S$ with $S^2 = +1$, just as those of
  the adjoint groups $SU(N)/Z_N$ or most of the exceptional Lie
  groups such as $G_2$.}        
Therefore, charge conjugation of the gauge fields in \qcd can be
undone by the constant $SU(2)$ gauge transformation $S =
\imag\sigma_2$. From now on we will use $T^a = \sigma^a/2$ for the color
generators, with 
\begin{equation}
{T^a}^T = {T^a}^* = - S T^a S^{-1} \; ,
\end{equation}
and reserve $\sigma_i$ ($\tau_i$) for the Pauli matrices in spinor
(flavor) space. Together with the charge conjugation matrix $C$ in
spinor space, likewise with $C^2= -1$, and complex conjugation denoted
by $K$ one then defines an antiunitary symmetry $T= SCK$ with $T^2 =
+1$ (in a real color representation with $S^2=+1$, one has $T^2=
-1$, correspondingly). This leads to the classification of the Dirac
operator by the Dyson index $\beta$ of random matrix theory
\cite{Kogut:1999iv,Kogut:2000ek}, with $\beta = 1$ for fermions in
the pseudoreal fundamental color representation of \qcd (or $\beta =4
$ in the real color representations of $SU(N)/Z_N$ or $G_2$).
  
Following \cite{Kogut:2000ek}, we start from the 
 standard kinetic part of the Euclidean \qcd Lagrangian, in the chiral basis,  
\begin{equation}
\mathcal{L}_\text{kin}= \bar{\psi}\slashed{D}\psi  =  \psi_L^\dagger \imag
\sigma_\mu D_\mu \psi_L - \psi_R^\dagger \imag \sigma_\mu^\dagger D_\mu
\psi_R \; , \label{eq:2}
\end{equation}
with hermitian $\gamma$-matrices, $\sigma_\mu = (-\imag, \vec\sigma
)$, and $\psi_{R/L}$, $\psi_{R/L}^*$ as independent Grassmann
variables with $\psi_{R/L}^\dagger\equiv \psi_{R/L}^{*\, T}$. 
The covariant derivative is $D_\mu=\partial_\mu+\imag A_\mu$, and the coupling is absorbed in the gauge fields $A_\mu=A_\mu^aT^a$.

The two terms in (\ref{eq:2}) get interchanged under the anti\-unitary
symmetry $T$. If we apply it only to the second term, say, by using
$(-\imag \sigma_2)$ for the chiral $R$-component of the charge
conjugation matrix $C$, {\it  i.e.}, changing variables to
$\tilde\psi_R = -\imag\sigma_2 S \psi_R^*$ and $\tilde\psi_R^* =
-\imag\sigma_2 S \psi_R$, we can therefore reexpress 
\begin{equation}
\mathcal{L}_\text{kin}=\Psi^\dagger\imag \sigma^\mu D_\mu \Psi \label{eq:3}
\end{equation}
in terms of the $2N_f$ $4$-dimensional spinors
$\Psi=(\psi_L,\tilde\psi_R)^T$ and
$\Psi^\dagger=(\psi_L^\dagger,\tilde\psi_R^\dagger)$. Because it is
now block diagonal, the $SU(2N_f)$ symmetry in the space combining
flavor and transformed chiral components is manifest in this form.
With the same transformation of variables the quarks' Dirac-mass term
becomes   
\begin{equation} 
m \bar{\psi}\psi = \smallfrac{m}{2} \big( \Psi^T \imag\sigma_2 S
\Sigma_0 \Psi \, -  {\Psi^*}^T\imag\sigma_2  S  \Sigma_0\Psi^*  \big)\, ,
\label{eq:4}
\end{equation}
where the symplectic matrix
\begin{equation} 
\Sigma_0=\begin{pmatrix}0& \mathds 1_{N_f}\\-\mathds 1_{N_f}&0\end{pmatrix}
\end{equation}
acts in the $2N_f$-dimensional extended flavor space. 
An explicit(dynamical) Dirac mass therefore explicitly(spontaneously)
breaks the original $SU(2N_f)$ down to the compact symplectic group
$Sp(N_f)$, sometimes also referred to as $USp(2N_f)$ reflecting the
fact that it is the intersection of the unitary $U(2N_f)$ and the symplectic
$Sp(2N_f,\mathds C)$, the invariance group of $\Sigma_0$ as bilinear form
on complex $2N_f$-vectors.   

For $N_f=2$ flavors the enlarged flavor symmetry group of \qcd is 
$SU(4)$, not $U(4)$ because of the axial anomaly, it replaces the
usual chiral and baryon number symmetries $SU(2)_L\times
SU(2)_R\times U(1)_B$. Just as this extended flavor $SU(4)$ shares its
15 dimensional Lie algebra with the group of rotations in 6
dimensions, $SO(6)$, its $Sp(2)$ subgroup leaving the Dirac-mass
term invariant has the 10 dimensional Lie algebra of $SO(5)$ (in fact
they are both the universal covers of the respective rotation
groups).  

Our brief review of the  \qcd symmetries so far holds for 
vanishing chemical potential. For $\mu\not= 0 $ but
$m=0$, the $SU(2N_f)$ symmetry is broken explicitly by $\mu
\bar\psi\gamma_0\psi $ to
$SU(N_f)_L\times SU(N_f)_R\times U(1)$. This is also easy
to see, from \Eqs{eq:2}, (\ref{eq:3}), as it amounts to introducing the term 
$\mu \bar\psi \gamma_0 \psi = \mu \Psi^\dagger B_0\Psi $ with
\cite{Kogut:2000ek}   
\begin{equation}
B_0 = -\gamma_0\Sigma_0 = \begin{pmatrix} \mathds 1_{N_f}&0 \\0& -\mathds 1_{N_f}
\end{pmatrix} \; .
\end{equation}
For $N_f=2$, in terms of the rotation groups, this symmetry breaking
pattern is locally the same as $SO(6) \to  SO(4)\times SO(2)$. 
  
When both $\mu$ and $m$ are non-zero, the unbroken flavor symmetry is
of course given by the common subgroup  $SU(2)_V \times U(1)$ of the two 
limiting cases $\mu\to 0$, $m\not=0$ or
$m\to 0$, $\mu\not=0$ discussed above. Whether, as an approximate
symmetry, it is more like the $Sp(2) \simeq SO(5)$ or like the 
$SU(2)_L\times SU(2)_R\times U(1)\simeq  SO(4)\times SO(2)$,
naturally depends on the relative sizes of Dirac mass $m$ and quark
chemical potential $\mu$.
 
More precisely, it is an exact result of chiral effective field
theory \cite{Kogut:1999iv,Kogut:2000ek}, that for baryon chemical
potential $\mu_B = 2\mu < m_\pi $ %$\propto \sqrt{m} $  
the approximate chiral symmetry breaking pattern remains that of $\mu = 0$ and
the vacuum alignment is $\langle \bar q q\rangle$-like with an
approximate $Sp(2) \simeq SO(5)$ if $m$ is sufficiently small, while
(at zero temperature) for $\mu_B = 2\mu > m_\pi $ a diquark condensate
develops and the vacuum  alignment starts rotating from being $\langle
\bar q q\rangle$-like to becoming more and more  $\langle q
q\rangle$-like as $\mu$ is further increased. The chiral condensate
then rapidly decreases, chiral symmetry appears to get restored but
the increasing chemical potential reduces it to the approximate
$SU(2)_L\times SU(2)_R \simeq SO(4)$ again. It is not the full flavor
symmetry of the $\mu\not=0$, $m=0$ case discussed above because we
have entered the diquark-condensation phase with spontaneous
baryon-number breaking, corresponding to superfluidity of the bosonic
baryons.     

Another exact result is that, at zero temperature and for $\mu_B <
m_\pi $, the onset of baryon condensation, the baryon density
remains zero and the thermodynamic observables must be independent of
$\mu$. Because this is far from obvious to verify explicitly in actual
calculations, it has been named the Silver Blaze Problem
\cite{Cohen:2003kd}. In order to be able to excite any states at zero
temperature, and with a gap in the spectrum, the relativistic chemical
potential needs to be increased beyond the mass gap in the
correlations to which it couples. Here, with a continuous
zero-temperature transition at $\mu_B = m_\pi$ this gap is simply
given by the baryon mass in vacuum which because of the extended
flavor symmetry in \qcd coincides with the pion mass, $m_B = m_\pi$.  
This latter property is of course special to $N_c=2$. The Silver Blaze
property will hold, as it does here, up to a quark chemical potential
of the order of $m_B/N_c$ (reduced by $1/N_c$ of the binding energy
per nucleon when the transition is of first order), in general, however.

\begin{figure}[ht]
\centering
\includegraphics[width=0.3\textwidth]{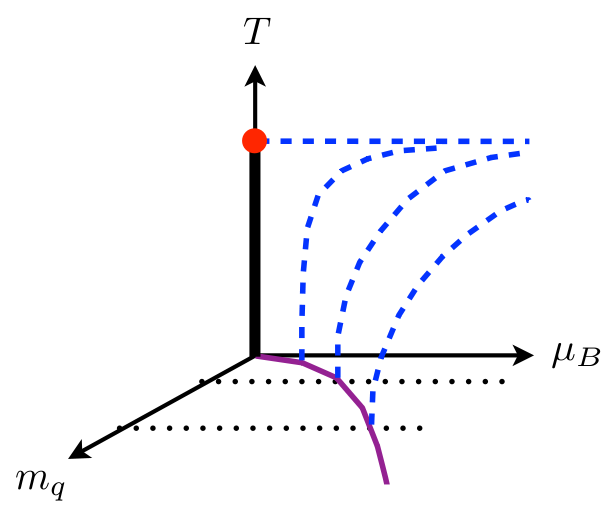}
\vspace{-.2cm}
\caption{Schematic phase diagram for \qcd  in the parameter space of
  temperature $T$, quark mass $m_q$ and baryon chemical potential
  $\mu_B$. 
\vspace{-.2cm} }
\label{fig:pdsketch}
\end{figure}

At finite temperature, a qualitative picture emerges for the phase diagram as
sketched in Fig.~\ref{fig:pdsketch}. The solid line in the $T=0$ plane
there represents the continuous zero-temperature %(quantum) phase 
transition with diquark condensation which is of mean-field
type. Because the quark mass $m_q$ scales quadratically with the pion
mass, it will occur along a parabola $m_q\propto \mu_B^2$. The thick
dashed lines represent the corresponding second-order transitions at finite
temperature in fixed $m_q$ planes of the $O(2)$ universality. The thick
line along the temperature axis is the magnetic first-order transition in the
 $\mu_B= 0$ plane which probably ends in a multi-critical point.
When viewed in the $\mu_B= 0$ plane, this is the critical endpoint in
the $O(6)$ universality class for the chiral phase transition in \qcd 
with its extended $SU(4)$ flavor symmetry. In the $m_q = 0$ plane, the
vacuum alignment will always be $\langle q q \rangle$-like, for
no-matter-how-small $\mu>0$. Therefore, in this plane one only has the
second-order $O(2)$ line which might end in the same point making it
multi-critical.

\subsection{Quark-meson-diquark model Lagrangian}
\label{sec:qmdLag}

The starting point of our model construction is the flavor structure
of the standard chiral condensate and the quark mass term which is
of the form $\Psi^T \Sigma_0 \Psi$. It therefore transforms under the
full flavor $SU(4)$ according to the six-dimensional antisymmetric
representation in the decomposition $4\otimes 4=10\oplus 6$.

The other components belonging to the same multiplet are obtained from
transformations 
\begin{equation}
\Psi\to U\Psi \; , \quad
 U =\exp(\imag \theta^a X^a) \, \in \, SU(4)/Sp(2)
 \; .
\end{equation}
Then, $\Psi^T \Sigma_0 \Psi \to \Psi^T \Sigma \Psi$, where, from
Cartan's immersion theorem, the whole coset $SU(4)/Sp(2) \cong S^5$ 
is obtained in this way via $\Sigma \equiv U^T \Sigma_0 U$.  The coset
elements $\Sigma$ are in turn parametrized by six-dimensional unit
vectors $\vec n$ as $\Sigma = \vec n \vec\Sigma$, with
$\Sigma_i^\dagger \Sigma_j + \Sigma_j^\dagger \Sigma_i= 2\delta_{ij}$
and $\vec \Sigma=(\Sigma_0,  \imag \Sigma_0 X^{a})$ such that 
$X^a$,  $a=1\dots 5$, form a basis for the coset generators
\cite{Brauner:2006dv}. Thus, one verifies explicitly that the
vector $\Psi^T \vec \Sigma \Psi $ transforms as a (complex)
six-dimensional vector under $SO(6)$.

A locally $SU(2)_c$ invariant linear sigma model Lagrangian can
therefore be defined by coupling the real $SO(6)$ vector of quark
bilinears $(\Psi^T\vec \Sigma\Psi+\text{h.c.})$ to a vector of mesonic
fields $\vec\phi=(\sigma,\vec{\pi},\text{Re}\,\Delta,\text{Im}\,\Delta)^T$
formed by the scalar $\sigma$ meson, the pseudoscalar pions $\vec\pi$
and the scalar diquark-antidiquark pair $\Delta$. This yields the
Lagrangian (now including color and spinor components again), 
\begin{equation}
\begin{split}
\label{eq:sigmamodel}
\mathcal{L}_\sigma=&\Psi^\dagger\imag \sigma^\mu
D_\mu\Psi+\frac{g}{2}(\Psi^T \imag \sigma_2
S\vec \Sigma\Psi - {\Psi^*}^T \imag \sigma_2
S\vec \Sigma\Psi^*)\vec \phi\\ 
&+\frac{1}{2} (\partial_\mu \vec \phi)^2 +V(\vec \phi),
\end{split}
\end{equation}
where $V(\vec \phi)$ is the meson and diquark potential whose precise
form will be specified later. A non-vanishing chemical potential
couples not only to the quarks but also to the bosonic
diquarks. Rewriting \Eq{eq:sigmamodel} in terms of the original
variables we obtain the quark-meson-diquark (QMD) model Lagrangian
\begin{equation}
\label{eq:sigmamodeloriginal}
\begin{split} \mathcal{L}_\text{QMD}=&\bar{\psi}\left(\slashed{D}+g(\sigma+\imag\gamma^5\vec{\pi}\vec{\tau})-\mu
    \gamma^0\right)\psi\\
  &+\frac{g}{2}\left(\Delta^* (\psi^T C \gamma^5\tau_2 S \psi)+
    \Delta(\psi^\dagger C \gamma^5\tau_2 S\psi^*)\right)\\
  &+\frac{1}{2} (\partial_\mu \sigma)^2+\frac{1}{2} (\partial_\mu \vec
  \pi)^2  +
  V(\vec \phi)\\
  &+\frac{1}{2} \big((\partial_\mu-2\mu \,
  \delta_{\mu}^{0})\Delta\big) 
 (\partial_\mu+2\mu\, \delta_\mu^0)\Delta^*\ ,
\end{split}
\raisetag{\baselineskip}
\end{equation}
with $C=\gamma^2\gamma^0$ and a flavor- and color-blind
Yukawa coupling $g$. With 
\begin{equation} 
 V(\vec\phi) = \frac{\lambda}{4} (\vec\phi^2 - v^2 )^2 - c\sigma \, ,
\label{eq:Vlinsig}
\end{equation} 
one obtains the corresponding $O(6)$ linear sigma model; and 
in the limit $\lambda\to\infty$, the bosonic part of $
\mathcal{L}_\text{QMD}$ is equivalent to the leading-order $\chi$PT
Lagrangian of Refs.~\cite{Kogut:2000ek}. To see this
explicitly it is best to start from the latter, use the
explicit coset parametrization of \cite{Brauner:2006dv} as given
above, and make the identifications $v= f_\pi= 2F$ and $c=f_\pi m_\pi^2
= 2Fm_\pi^2$. It maybe worthwhile mentioning that the coefficient of
the leading term in $\mu$ of the $\chi$PT Lagrangian,
$\mu^2\mbox{tr}(\Sigma B^T\Sigma^\dagger B) $ with $B =
UB_0U^\dagger$, which was fixed from gauging the flavor $SU(4)$ in
\cite{Kogut:1999iv}, here simply follows from $-2\mu^2 |\Delta|^2$ as
part of the kinetic term of the complex scalar diquark field $\Delta$
with chemical potential $\mu_B=2\mu$. This implies in particular, that
the meson/diquark potential $V(\vec\phi)$ itself, up to the explicit breaking by
$-c\sigma$, which needs to be only $SO(4) \times SO(2)$ invariant in
general at finite $\mu$, must remain $SO(6)$ invariant, however, at
this leading order, $\mathcal O(\mu^2)$, and therefore at $\mathcal
O(\phi^2)$ in the fields, likewise.  We can thus only
have an $SO(6)$ invariant mass term in $V(\vec\phi)$.   

In the following it will be more convenient to rewrite the Lagrangian
in terms of Nambu-Gorkov-like spinors
$\Psi=\left(\begin{smallmatrix}\psi_r\\
    \tau_2\psi_g^C \end{smallmatrix}\right)$, where $\psi_r$
($\psi_g$) denote the red (green) color components of $\psi$ and
$\psi^C\equiv C\bar{\psi}^T$ as in \cite{Brauner:2009gu}. This yields 
\begin{equation}
\begin{split}
\label{eq:LagrangianNG}
\mathcal{L}_\text{QMD}=&\bar{\Psi} S_0^{-1}\Psi+\frac{1}{2} (\partial_\mu \sigma)^2+\frac{1}{2} (\partial_\mu \vec
  \pi)^2 + V(\vec\phi)\\
&+\frac{1}{2} \big( (\partial_\mu-2\mu \delta_{\mu}^0) 
    \Delta\big)(\partial_\mu+2\mu \delta_\mu^0)\Delta^*,
\end{split}
\end{equation}
where
\begin{equation}
\label{eq:S0-1}
S_0^{-1}=\left(\begin{smallmatrix}
    \slashed{\partial}+g(\sigma+\imag\gamma^5\vec{\pi}\vec{\tau})-\gamma^0\mu
    &  g\gamma^5 \Delta \\
 - g\gamma^5\Delta^*
    &  \slashed{\partial}+g(\sigma-\imag\gamma^5\vec{\pi}\vec{\tau})+\gamma^0\mu \end{smallmatrix}\right).
\end{equation}
%%%%%%%%%%%%%%%%%%%%%%%%%%%

Gauge field dynamics and confinement effects can be modeled 
also in \qcd by including a constant Polyakov-loop variable as a background
field as in the NJL model \cite{Brauner:2009gu}, and analogous to what
is commonly done in the so-called Polyakov-loop-extended quark-meson
models of three-color QCD \cite{Schaefer:2007pw,Schaefer:2009ui,  
  Herbst:2010rf}. To this end one introduces a constant temporal
background gauge field $A_\mu=A_0\delta_{\mu 0}$ which is furthermore
assumed to be in the Cartan subalgebra as in the Polyakov gauge, {\it
  i.e.}, for $SU(2)_c$ simply given by $A_0=T^3 2a_0$. This leads to the
Polyakov loop variable 
\begin{equation}
\label{eq:polyvar}
\Phi\equiv\frac{1}{2}\Tr_c e^{\imag \beta A_0}=\cos(\beta a_0),
\end{equation}
to model a thermal expectation value of the color-traced Polyakov loop 
at an inverse temperature $\beta = 1/T$, as an order parameter for the
deconfinement transition at vanishing chemical potential.  
The covariant derivative
$D_\mu=\partial_\mu-\imag \delta_{\mu 0} A_0$ leads to an additional
contribution of the form $-\imag\bar{\psi}\gamma^0 T^3 2a_0 \psi$
which can be rewritten as $- \imag\bar{\Psi}\gamma^0 a_0 \Psi $ in
terms of the spinor field $\Psi$.  Finally, we then arrive at the
Polyakov-loop-extended quark-meson-diquark model (PQMD) Lagrangian,
\begin{equation}
\label{eq:LagrangianPQM}
\mathcal{L}_\text{PQMD}=\mathcal{L}_\text{QMD}- \imag \bar{\Psi} 
\left(\begin{smallmatrix} \displaystyle \gamma^0 a_0 &
    \displaystyle 0\\
\displaystyle 0 & \displaystyle
\gamma^0 a_0 \end{smallmatrix}\right)\Psi
+\mathcal{U}_\text{Pol}(\Phi),   
\end{equation}
with $\mathcal{L}_\text{QMD}$ defined in \Eq{eq:LagrangianNG} and
$\mathcal{U}_\text{Pol}(\Phi)$ is the Polyakov-loop potential
\cite{Brauner:2009gu} which is commonly fitted to lattice results, but
which can also be computed with functional methods
\cite{Braun:2007bx,Marhauser:2008fz}. In contrast to the three-color case the 
Polyakov-loop potential is a function of one single real variable
$\Phi$ here, even in the presence of a diquark condensate.

%%%%%%%%%%%%%%%%%%%%%%%%%%%%%%%%%%%%%%%%%%%%%%%%%%%%%%%%%%%%%%%%%%%%%%%%%%%%%
% Mean field section
%%%%%%%%%%%%%%%%%%%%%%%%%%%%%%%%%%%%%%%%%%%%%%%%%%%%%%%%%%%%%%%%%%%%%%%%%%%%%
\section{Mean-Field Thermodynamics\label{sec:MF}}

The grand potential in mean-field approximation is obtained by
integrating over the quark fields and neglecting bosonic
fluctuations. This means that all mesonic and diquark fields are
replaced by their constant expectation values $\sigma\equiv \langle
\sigma\rangle$, $\Delta\equiv \langle \Delta\rangle$, $\Delta^*\equiv
\langle \Delta^*\rangle$ and $\vec{\pi}\equiv \langle
\vec{\pi}\rangle=\vec{0}$. In momentum space we then obtain,
\begin{equation}
\label{eq:LagrangianMF}
\mathcal{L}_{\text{PQMD}} %\sigma+\mu+\Phi
  ^{\text{MF}}=\bar{\Psi}\left(
  S_{0,\text{MF}}^{-1}-\imag\gamma^0 a_0 \right)\Psi+V_\text{MF}(\sigma, d^2)+\mathcal{U}_\text{Pol}(\Phi), 
\end{equation}
where 
\begin{equation}
S_{0,\text{MF}}^{-1}=\left(\begin{smallmatrix}
    -\imag\slashed{p}-\gamma^0\mu+g \sigma
    &  g\gamma^5 \Delta \\
 - g\gamma^5\Delta^*
    &  -\imag\slashed{p}+\gamma^0\mu+g \sigma \end{smallmatrix}\right)\, ,
\end{equation}
and $V_\text{MF}(\sigma, d^2)=
({\lambda}/4)\left(\sigma^2+d^2 -v^2\right)^2 -c\sigma -2\mu^2 d^2$
with $d^2 \equiv |\Delta|^2$ is the bosonic effective potential. The
last term comes from the kinetic     
diquark part of \Eq{eq:LagrangianNG} and is included in the effective
potential here. This term and the explicit chiral symmetry
breaking by $-c\sigma$ break the $SU(4)$ symmetry of the
effective potential $V_\text{MF}$. The details of the parameter fixing
and the values used in the numerical calculations are given in
Appendix~\ref{sec:paramfixing}.

The fermion-loop integration then yields for
the grand potential $\Omega$,
\begin{equation}
\begin{split}
\Omega(T,\mu)=&-T\sum_{n\in Z}\int \frac{\d^3p}{(2\pi)^3}\Tr \log
\big(S_{0,\text{MF}}^{-1}-\imag \gamma^0a_0\big)\\
&+V_\text{MF}(\sigma, d^2)+\mathcal{U}_\text{Pol}(\Phi),
\end{split}
\end{equation}
where the trace runs over internal indices (Dirac-, flavor- and
Nambu-Gorkov space) and we
sum over antiperiodic Matsubara modes $\nu_n=(2n+1)\pi T$. The
four distinct eigenvalues of $\gamma^0 S_{0,\text{MF}}^{-1}$ are given by $\pm
E_{p}^{+}-\imag \nu_n$ and $\pm E_{p}^{-}-\imag \nu_n$ with
\begin{equation}
\begin{split}
\label{eq:MFeigenvalues}
E_{p}^{\pm}&=\sqrt{g^2d^2 +
  {\epsilon^\pm_{p}}^2}\,,\\
\epsilon^\pm_{p}&=\epsilon_{p}\pm\mu\quad\text{and}\quad\epsilon_{p}=\sqrt{
  \vec{p}^2 + g^2\sigma^2}. 
\end{split}
\end{equation}
The Matsubara sum can be performed analytically with the result
\begin{equation}
\label{eq:MFgcpotential}
\begin{split}
  \Omega( \sigma, d^2, \Phi)=&-4\int \frac{\d^3p}{(2\pi)^3}\Bigl\{E_{p}^+ + E_{p}^- \\
  &+T \log\left(1+2\Phi e^{-\beta E_{p}^{+}}+e^{-2\beta
      E_{p}^{+}}\right)\\&+T\log\left(1+2\Phi e^{-\beta
      E_{p}^{-}}+e^{-2\beta
     E_{p}^{-}}\right)\Bigr\}\\&+V_\text{MF}(\sigma, d^2)+\mathcal{U}_\text{Pol}(\Phi)\,.
\end{split}
\end{equation}
When the bosonic potential  $V_\text{MF}$
is replaced by $ M^2 (\sigma^2 + d^2) - c\sigma$, with $M^2 = g^2/(4G)
$ and $c = 2gm_0 /(4G)$, this  
coincides with the Hubbard-Stratonovich transformed PNJL model
result \cite{Brauner:2009gu} with four-quark coupling $G$ and
current-mass parameter $m_0$. Note that the model
independent $-2\mu^2d^2$ term from chiral effective field theory, which
is included in the bosonic part of the (P)QMD model, does not explicitly
show up in the grand potential of the (P)NJL model.  Minimizing the
thermodynamic potential with respect to the constant mean fields
$\sigma,d,\Phi$ 
leads to the gap equations,
\begin{equation}
\label{eq:mfgap}
\frac{\partial \Omega}{\partial \sigma}=\frac{\partial
  \Omega}{\partial d}=\frac{\partial \Omega}{\partial \Phi}=0\ ,
\end{equation}
whose simultaneous solution yields the temperature and chemical
potential dependent condensates $\sigma$, $d$ and $\Phi$.

%%%%%%%%%%%%%%%%%%%%%%%%%%%%%%%%%%%%%%%%%%%%%%%%%%%%%%%%%%%%%%%%%%%%%%%%%%%%%
\subsection{Vacuum contributions}
\label{sec:MFvac}

The fermion-loop contribution to the grand potential in mean-field
approximation, \Eq{eq:MFgcpotential}, contains an ultraviolet
divergent vacuum part. In the standard {\it no-sea} mean-field
approximation one usually dismisses this vacuum contribution
to the bulk thermodynamics. For some phenomenological consequences of
this additional approximation and its influence on mean-field results,
see Ref.~\cite{Skokov:2010sf,Schaefer:2011ex} and the references therein. Here we add
an observation concerning this mean-field ambiguity of the quark-meson
model when viewed as the $d\to 0$ limit of the quark-meson-diquark
model grand potential. Because the Polyakov-loop contributions are
irrelevant here we set $\Phi=1$. 

\begin{figure}[ht]
\centering
\includegraphics[width=0.47\textwidth]{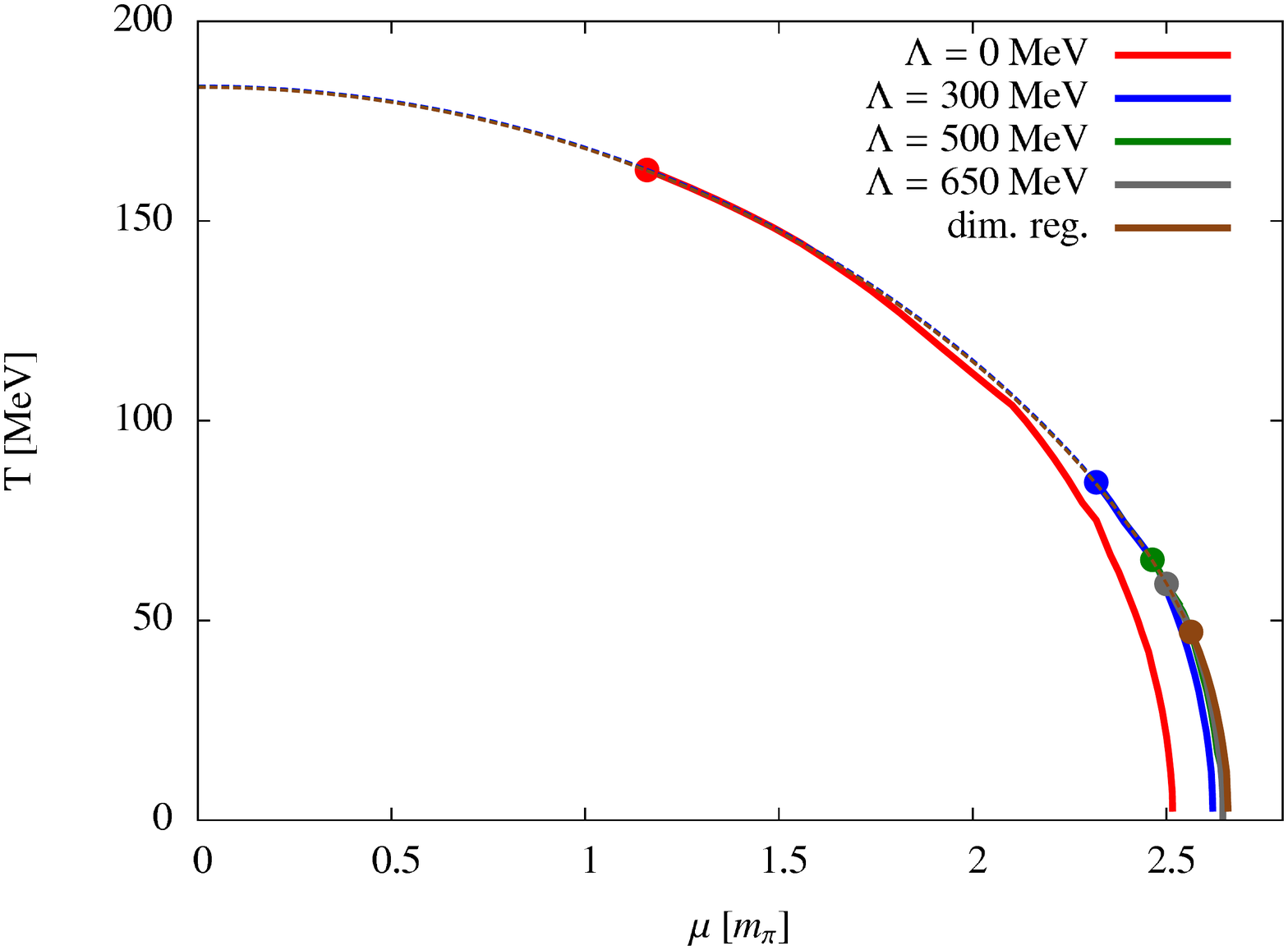}
\vspace{-.2cm}
\caption{Standard $N_c=2$ QM phase diagram: dependence of the location
  of the CEP on the vacuum-term cutoff $\Lambda$ in comparison to
  dimensional regularization. 
\vspace{-.5cm} }
\label{fig:pdd0old}
\end{figure}

For $d=0$ \Eq{eq:MFgcpotential} superficially appears to reduce to the
conventional mean-field expression for the grand potential of the
quark-meson model \cite{Scavenius:2000qd} up to an overall $N_c$ in
front of the fermion-loop contribution $\Omega_q$, 
\begin{equation}
\label{eq:MFfermpress}
  \Omega_q = \Omega_q^{\mathrm{vac}}  -4 N_c T\int\frac{\d^3p}{(2\pi)^3}
  \sum_\pm \log\left(1+e^{-\beta (\epsilon_{p}\pm \mu)}\right).
\end{equation}
We illustrate the effect of the vacuum contribution
for $N_c=2$ colors but without diquark condensate ({\it i.e.} with
$d=0$) in Fig.~\ref{fig:pdd0old}.  To regularize the vacuum term
$\Omega^{\mathrm{vac}}_q$  we employ a simple three-momentum cutoff
$\Lambda$ and assess the dependence of the phase structure on
$\Lambda$. The parameters are fixed to reproduce an $\sqrt{N_c}$-scaled 
$f_\pi=76$ MeV and a pion screening mass of $m_\pi=138$ MeV. In each
case the sigma meson mass is adjusted so as to yield a common value
for a chiral transition at $\mu=0$ of $T_c\approx 183$
MeV. $\Lambda=0$ corresponds to the no-sea approximation.
The dependence of the position of the critical endpoint (CEP) at
$\mu_c$ on the cutoff $\Lambda$ is clearly visible in
Fig.~\ref{fig:pdd0old}. With increasing $\Lambda$ its location gets shifted
towards larger chemical potentials and approaches the dimensionally
regularized result  \cite{Skokov:2010sf} when $\Lambda/\mu_c $ is
sufficiently large. 

More carefully, however, one observes that the fermion-loop
in the no-sea approximation ($\Omega_q^\mathrm{vac} = 0$
when $\Lambda=0$) in \Eq{eq:MFfermpress} does not tend
to zero for $T\to 0$ when  $\mu > g\sigma $, but still contains
temperature independent contributions from momenta with $\vec p^2 <
\mu^2-g^2\sigma^2$.  

On the other hand, the $d\to 0$ limit of \Eq{eq:MFgcpotential} with 
\begin{equation}
E^-_{p}  \to  | \epsilon_{p}-\mu| 
\end{equation}
yields a grand potential of the quark-meson model
for two colors which depends only on the chiral condensate
$\sigma$ but which differs from the conventional expression
by the appearance of the modulus of the quasi-particle energies,
%$|\epsilon_{p}-\mu|$,
\begin{equation}
\label{eq:MFfermpressmod}
\begin{split}
  \Omega(\sigma)=&-4\int
  \frac{\d^3p}{(2\pi)^3}\Bigl\{\epsilon_{p}+\mu +
  |\epsilon_{p}-\mu| \\  
  &\quad+2 T \log\left(1+\,e^{-\beta(
      \epsilon_{p}+\mu)}\right)\\&\quad+ 2 T\log\left(1+\,e^{-\beta|
      \epsilon_{p}-\mu|}\right)\Bigr\}+V_\text{MF}(\sigma,0)\,.
\end{split}
\end{equation}
The last two terms herein deserve to be called thermal now, as they do
vanish at zero temperature for all $\mu$. 
When vacuum and the thermal contributions are regularized in the
same way we can recover the usual expression by means of the identity
\begin{equation}
|x| + 2 \log(1+\exp(-|x|)) %= \log(\cosh(|x|/2)) 
= 2\log\cosh(x/2) + 2\log 2  \, . 
\end{equation}
This is for example the case in the NJL model if one chooses to
regulate both thermal and vacuum parts with a three-momentum cutoff,
but this is not what is usually done in the quark-meson model where
the ultraviolet finite thermal contributions are meant to be fully
retained. The cutoff in the phase diagrams of Fig.~\ref{fig:pdd0old}
was applied only to $\Omega_q^\mathrm{vac}$ in \Eq{eq:MFfermpress},
likewise. Otherwise the picture would change yet again. This is a bit
of a grain of salt for the no-sea mean-field approximation in
quark-meson models which is best motivated phenomenologically as
modelling the restoration of chiral symmetry at $T=0$ for large
chemical potentials. Luckily, the problem is irrelevant altogether,
once fluctuations are included via the functional renormalization
group for which the quark-meson model shows its true uses.

Meanwhile, for the mean-field analysis of our quark-meson-diquark
model including the possibility of diquark condensation with
$d\not=0$, we really have no option other than the perhaps anyway more
natural splitting of thermal and vacuum contributions based on 
the modulus of the quasi-particle energy as the $d\to 0$ limit
of $E^-_{p}$ in the  QMD model mean-field grand potential,
\Eq{eq:MFgcpotential}.    

If one considers the difference between cutting off the vacuum term 
in \Eq{eq:MFfermpress} as compared to the one in 
\Eq{eq:MFfermpressmod} as measure for the reliability of the
calculation, one is led to conclude that the cutoff  $\Lambda$ in
$\Omega_q^\mathrm{vac}$ should always be larger than the chemical
potential $\mu$.  

In the following we will continue to regulate vacuum terms with a
sharp momentum cutoff mainly because dimensional regularization, as
applied to the three-color PQM model in \cite{Skokov:2010sf}, due to
the structure of $E_{p}^\pm$, gets too complicated for a semi-analytic
treatment here, with full diquark mean fields from the grand potential
in \Eq{eq:MFgcpotential}.

%%%%%%%%%%%%%%%%%%%%%%%%%%%%%%%%%%%%%%%%%%%%%%%%%%%%%%%%%%%%%%%%%%%%%%%%%%%%%
\subsection{Diquark condensation}
\label{sec:DiqCond}

Independent of the discussion in the previous section and of the
influence of fluctuations, we know for two-color QCD that 
the quark-meson-model-like phase diagrams of the form as those in 
Fig.~\ref{fig:pdd0old} are wrong. The exact chiral effective field
theory results \cite{Kogut:1999iv,Kogut:2000ek} from the symmetries
and breaking patterns as reviewed in Sec.~\ref{sec:chsymbreaking} tell
us that we must include the diquark condensate along with the chiral
condensate and base our mean-field analysis on
Eqs.~(\ref{eq:MFgcpotential}), (\ref{eq:mfgap}) in order to describe
the superfluid diquark phase starting at a critical line $\mu_c(T)$
with $\mu_c(0) = m_\pi/2 $ (or $\mu_B = m_B$).  

\begin{figure}[ht]
\centering
\includegraphics[width=0.47\textwidth]{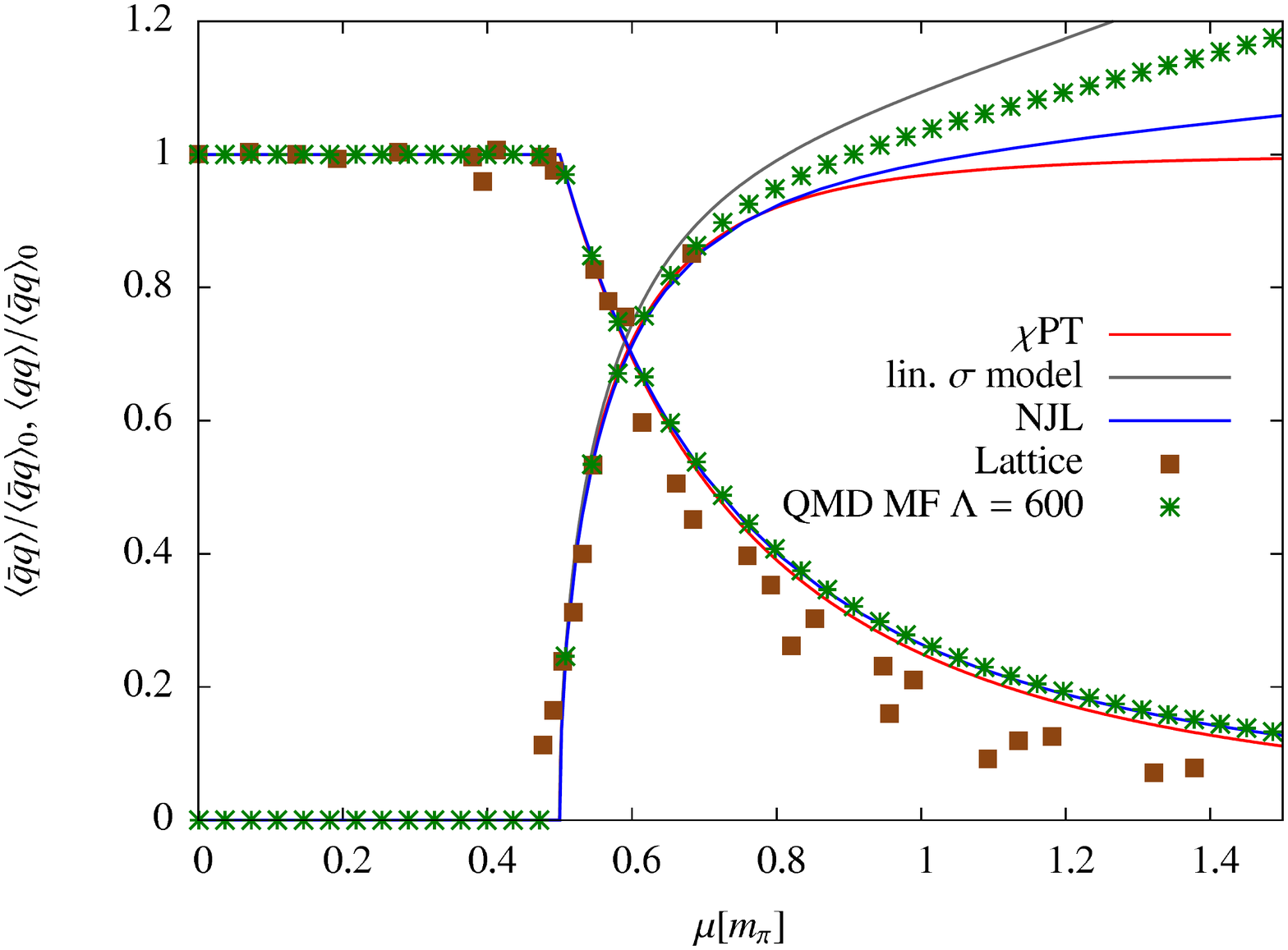}
\vspace{-.2cm}
\caption{Condensates at $T=0$ (NJL parameter values from
  \cite{Ratti:2004ra}; lattice data from \cite{Hands:2000ei}; linear
  sigma model with $m_\pi=138$ MeV and $m_\sigma=680$ MeV). 
\vspace{-.2cm} }
\label{fig:t0axismf}
\end{figure}

The resulting chemical potential dependence of the chiral and diquark
condensates at zero temperature is shown in Fig.~\ref{fig:t0axismf}
where we compare the prediction from leading order chiral perturbation
theory \cite{Kogut:2000ek}, the NJL \cite{Ratti:2004ra} and the
linear sigma model \cite{Andersen:2010vu} results with lattice data
\cite{Hands:2000ei} and our quark-meson-diquark model mean-field
result with vacuum contribution from  Eqs.~(\ref{eq:MFgcpotential}),
(\ref{eq:mfgap}). 

The $T=0$ onset of diquark condensation at $\mu_c(0)={m_\pi}/{2}$  
as an exact result is built-in in $\chi$PT and the $O(6)$ linear sigma
model as discussed in Sec.~\ref{sec:qmdLag}. Therefore, it also holds
for the screening mass of the pion from the bosonic potential in our
adapted no-sea approximation, which reduces to the linear sigma
model at $T=0$ by definition. They both go beyond the 
leading order chiral perturbation theory in that they include effects of a 
finite sigma meson mass. The linear sigma model expressions for
the $T=0$ condensates are \cite{Andersen:2010vu}
\begin{equation}
\label{eq:nonlinsigmamodelcond}
\begin{split}
\frac{\sigma}{\sigma_0}&=\left\{
  \begin{array}{l l}
    1 & \quad \text{for $\mu<\mu_c$}\\
    \frac{1}{x^2} & \quad \text{for $\mu>\mu_c$}\\
  \end{array} \right.\\
  \frac{|\Delta|}{\sigma_0}&=\left\{
  \begin{array}{l l}
    0 & \quad \text{for $\mu<\mu_c$}\\
    \sqrt{1-\frac{1}{x^4}+2\frac{x^2-1}{y^2-1}} & \quad \text{for $\mu>\mu_c$}\\
  \end{array} \right.,
  \end{split}
\end{equation} 
where $x=2\mu/m_\pi$ and $y=m_{\sigma}/m_{\pi}$. The only difference
between these and the $\chi$PT result  \cite{Kogut:2000ek} is the
$y$-dependent term in the diquark condensate which reduces to the   
 $\chi$PT formula for $y \to\infty $. Note that the chiral condensate 
does not depend on the sigma meson mass whereas the diquark condensate
does via $y$ which explains the variations of the diquark
condensates at large $\mu$ in Fig.~\ref{fig:t0axismf}. 

Beyond the no-sea approximation, one
needs to distinguish between screening and pole mass. Only the
latter agrees with the onset baryon chemical potential at the 
mean-field level, in general. We will discuss this in more detail in
the next subsection. 

The overall agreement of our $T=0$ results with the existing
literature in  Fig.~\ref{fig:t0axismf} is very reassuring. In
particular, there is no dependence on  
the chemical potential for $\mu<\mu_c$ in accordance with the Silver
Blaze property as also discussed more in  Sec.~\ref{sec:polemass} below.

\begin{figure}[ht]
\centering
\includegraphics[width=0.48\textwidth]{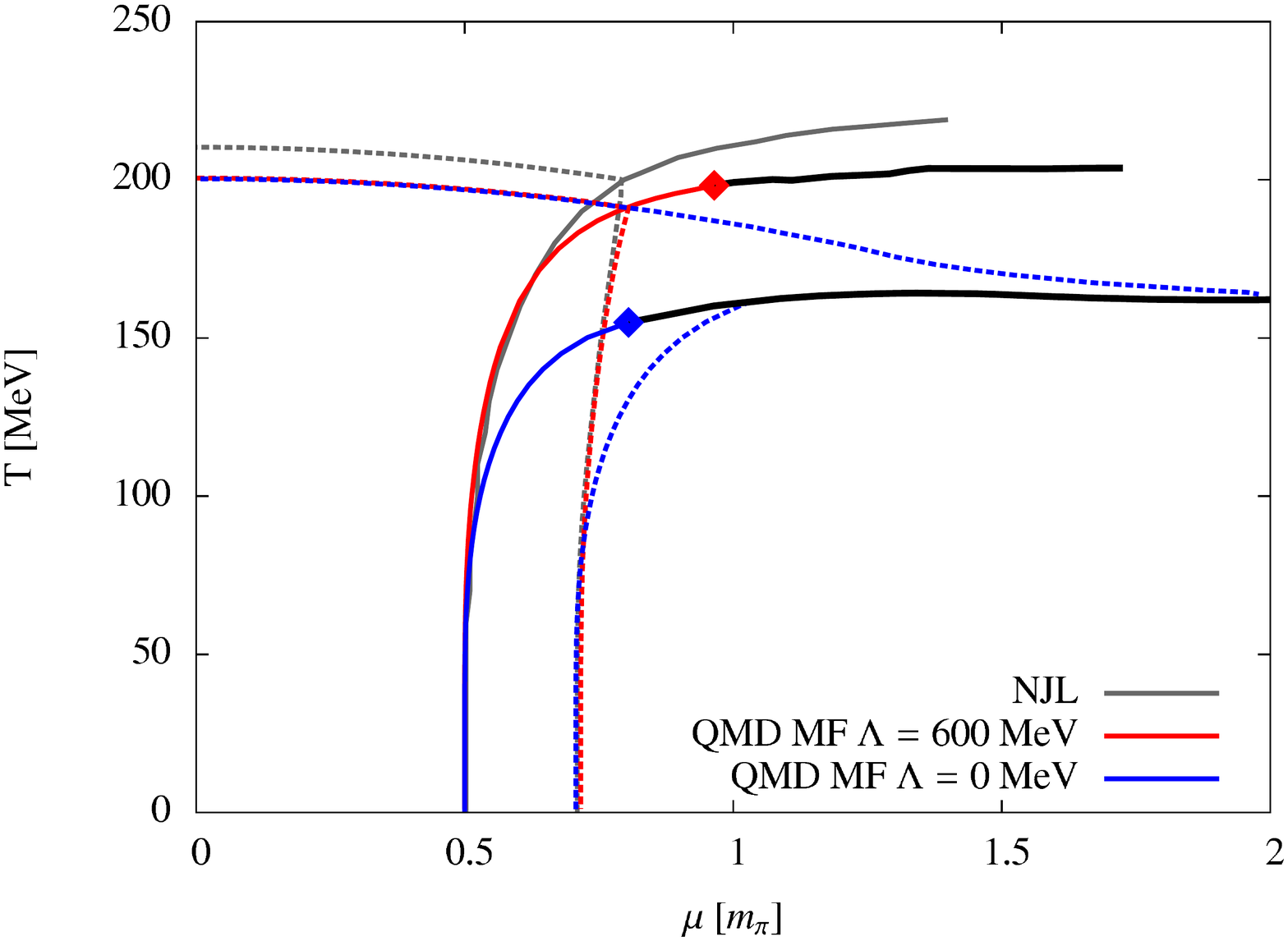}
\vspace{-.2cm}
\caption{Mean field phase diagrams: QMD model (with and without vacuum
  term) vs. NJL model (with parameters from \cite{Ratti:2004ra});
  dashed: chiral crossover; solid: second order transition; thick
  solid black: first order transition; diamonds: tricritical points as
  predicted in \cite{Splittorff:2001fy}.  
\vspace{-.2cm} }
\label{fig:pdmf}
\end{figure}

Common to all studies, the chiral condensate decreases with increasing
$\mu$ in the diquark condensation phase above $\mu_c$. As a result,
the quark-meson-model-like $1^\mathrm{st}$ order transition line and
CEP at $\mu$ around $2.5 m_\pi$ are completely gone, as seen also in 
Fig.~\ref{fig:pdmf}. 

Fig.~\ref{fig:pdmf} shows the phase diagram of two QMD model mean-field
calculations in comparison with the result of an NJL model recalculation
following Ref.~\cite{Ratti:2004ra}. For small chemical
potentials and small temperatures one finds the chirally broken phase
with vanishing diquark condensate crossing over with increasing
temperature to the phase with asymptotically restored chiral symmetry
as usual. In addition, there is the diquark condensation
phase for $\mu > \mu_c(T) $ characterized by non-zero baryon
density.  

The NJL model calculations show a continuous diquark condensation
transition throughout the whole phase diagram \cite{Brauner:2009gu}.
In contrast, in our mean-field QMD model results here, the second
order line $\mu_c(T)$ ends in a tricritical point from where on
towards larger chemical potentials it becomes a $1^\mathrm{st}$ order
transition. Such a behavior was also predicted at next-to-leading
order $\chi$PT \cite{Splittorff:2001fy}, where it was concluded that 
this tricritical point occurred at $\mu_c\approx 0.57  m_\pi$ and
$T \approx 220$ MeV. The temperature range is quite comparable here,
but its precise location depends on the value for the vacuum-term cutoff
$\Lambda$ and moves towards larger chemical potentials with increasing
$\Lambda$. Moreover, as we will see below, the first-order transition
will be washed out by fluctuations which will make it second-order
again, and it is hence a mean-field artifact.

%%%%%%%%%%%%%%%%%%%%%%%%%%%%%%%%%%%%%%%%%%%%%%%%%%%%%%%%%%%%%%%%%%%%%%%%%%%%%
\subsection{Pole versus screening masses\label{sec:polemass}}

The definition of meson/diquark masses is absolutely crucial already
at mean-field level to get results which are consistent with the
Silver Blaze property. As soon as one goes beyond the no sea
approximation there is an important distinction between the screening
mass at zero external momentum, which is just determined by the
curvature of the effective potential, and the pole mass which takes
into account a non-vanishing external momentum. The latter is the
natural choice in NJL model calculations but has so far not been taken
into account in QM model studies. Both definitions coincide for
massless particles and in the no sea approximation.

The meson/diquark pole masses are defined as the zeroes of the determinant 
of their inverse propagator,
\begin{equation} 
\Gamma^{(2)}_{ij}(p)=\Gamma^{(2)}_{\text{tl}}(p)_{ij}+\Pi_{ij}(p) \, ,
\end{equation}
where, for the pions,
\begin{equation} 
%\Gamma^{(2)}_{\text{tl}}(p)_{11} =& p^2 - m^2 + 3 \lambda\phi^2
\Gamma^{(2)}_{\text{tl}}(p)_{ij} = \big( p^2 - m^2 + \lambda \phi^2
\big)  \delta_{ij} \; , \;\; i,j = 2,3,4\, , %\nonumber  
\label{eq:tlG2pions}
\end{equation}
with $m^2 = \lambda v^2$ from the tree-level potential
(\ref{eq:Vlinsig}) and $\phi^2 = \sigma^2+d^2$. 
When the diquark condensate $d$ is non-zero, we choose it to lie in the
$\phi_5 = \text{Re}\,\Delta$ direction without loss, the sigma meson
mixes with the scalar diquark pair already at tree-level 
through the $O(6)$ linear sigma model potential (\ref{eq:Vlinsig}). 
In the 3-dimensional subspace of sigma
%$\phi_1 = \sigma $ 
and diquarks in the real basis $\phi_5 =
\text{Re}\,\Delta$, $\phi_6 = \text{Im}\,\Delta$, 
\begin{widetext}
\begin{equation}  
%\begin{split}
\Gamma^{(2)}_{\text{tl}}(p) =  
\begin{pmatrix} 
p^2 - m^2  + \lambda\phi^2  + 2\lambda \sigma^2 &  2 \lambda \sigma d & 0 \\
2 \lambda\sigma d &   p^2 - 4\mu^2 - m^2 + \lambda\phi^2  + 2\lambda
d^2 & -4 \mu p_0 \\ 
0& 4\mu p_0 &  p^2 - 4\mu^2  - m^2 + \lambda\phi^2   
\end{pmatrix} \, .
%\end{split} %\nonumber
\label{eq:tlG2sigDelta}
\end{equation}
\end{widetext}
The RPA polarization functions are obtained from evaluating the
fermion-loop integrals with external momentum $p$ in the usual way, 
\begin{equation}
\label{eq:rpadefinition}
\Pi_{ij}(p)=\Tr_q\left[ \frac{\partial \Gamma^{(2)}_F}{\partial \phi_i}\Big|_{\phi_{\mathrm{MF}}}\hspace{-1mm} G_{\mathrm{MF}}(p+q) \frac{\partial \Gamma^{(2)}_F}{\partial \phi_j}\Big|_{\phi_{\mathrm{MF}}}\hspace{-1mm} G_{\mathrm{MF}}(q)\right],
\end{equation}
where
$G_{\mathrm{MF}}={\left(\Gamma^{(2)}_F|_{\phi_{\mathrm{MF}}}\right)}^{-1}$. 
They can be found in the NJL model literature
\cite{He:2005nk,Xiong:2009zz,Brauner:2009gu}, originally from the
two-flavor three-color standard NJL model with isospin chemical
potential and pion condensation. Since the available expressions are
either incomplete or at variance with our computations, we have
recomputed them and list the complete explicit expressions for
these polarization functions as a convenience to the reader 
in Appendix~\ref{sec:rpamass}.   

To find the pole masses we use $p = (-\imag \omega,\vec0)$ and the
somewhat sloppy notations $\Gamma^{(2)}(\omega ) \equiv
\Gamma^{(2)}(p=(-\imag \omega,\vec0))$, $\Pi(\omega ) \equiv
\Pi(p=(-\imag \omega,\vec0))$, to solve 
\begin{equation}
\det \Gamma^{(2)} (\omega) =
0 \; , \;\;  \mbox{for} \;\; \omega = m_k\; ,\;\;  k=1,\dots 6 \, .
\end{equation}
The polarization integrals are ultraviolet divergent. As before, we
use a spatial momentum cutoff $\Lambda $ for the
temperature-independent contributions. Making the $T$-dependence
explicit, we may thus write, 
\begin{equation} 
\begin{split}
  \Pi^\mathrm{reg}(\omega,T) =& \; \Pi^\mathrm{th}(\omega,T) +
  \Pi^\mathrm{vac}_\Lambda(\omega) \; , \;\; \mbox{where}\\
  \Pi^\mathrm{th}(\omega,T) =&  \; \Pi(\omega,T) - \Pi(\omega,0)  
\end{split}
\end{equation}  
is ultraviolet finite. In the normal phase with $d=0$, {\it i.e.}, for
$\mu $ below the onset of diquark condensation at $\mu_c(T)$, the
polarization integrals are diagonal in the basis where $\Delta =
\phi_5 + i\phi_6$ and   $\Delta^* = \phi_5 - i\phi_6$. 
The polarization matrix $\Pi(\omega,T) $ from App.~\ref{sec:rpamass}
is then diagonal with entries \cite{Brauner:2009gu},
\begin{equation}
\begin{split}
\Pi_\sigma(\omega,T) =&  16 N_c g^2\!\!\int\!\! \frac{\d^3q}{(2\pi)^3} \frac{ \vec q^2}{\epsilon_{q}}\, \frac{1-N_q(\epsilon_q^-)-N_q(\epsilon_q^+)}{\omega^2-4 \epsilon_{q}^2}\\
&+ 4 N_c g^2\delta_{\omega,0}\!\!\int\!\! \frac{\d^3q}{(2\pi)^3} \frac{ g^2\sigma^2}{\epsilon_{q}^2}\left(N_q'(\epsilon_q^+)+N_q'(\epsilon_q^-)\right)\\
\Pi_\pi(\omega,T) =&  16 N_c g^2 \!\!\int\!\! \frac{\d^3q}{(2\pi)^3}
\frac{\epsilon_{q}(1-N_q(\epsilon_q^-)-N_q(\epsilon_q^+))}{\omega^2-4 \epsilon_{q}^2}\\
\Pi_\pm(\omega,T) =&  4 N_c g^2\!\! \int\!\! \frac{\d^3q}{(2\pi)^3}
\Bigg(\frac{1-2 N_q(\epsilon_q^\mp)}{\omega-2\epsilon_q^\mp}\!-\!\frac{1-2 N_q(\epsilon_q^\pm)}{\omega+2\epsilon_q^\pm} \Bigg)
\end{split}  \label{eqs:polnormal}
\end{equation}
with $\epsilon^\pm_{q} = \sqrt{\vec{q}^2 + g^2\sigma^2}\pm \mu$,
$\Pi_\pm$ for $\Delta$, $\Delta^*$ and Polyakov loop enhanced
quark/antiquark occupation numbers
\begin{equation}
\label{eq:PLenhancedoccnumbers}
N_q(E)\equiv N_q(E;T,\Phi)=\frac{1+\Phi e^{\frac{E}{T}}}{1+2\Phi e^{\frac{E}{T}} +e^{\frac{2E}{T}}},
\end{equation}
which simplify to the Fermi-Dirac distribution for $\Phi=1$.

As usual, these expressions are obtained from analytically continuing
the results for imaginary discrete values $\omega = \imag 2\pi T n$
corresponding to the Matsubara frequencies in imaginary time.  
To make the continuation unique, one usually assumes in addition that  
the polarization functions are well behaved at complex infinity with cuts only
along the real axis. Then, it follows that the correspondingly
continued expressions for finite spatial momenta, $\Pi(p=(-\imag
\omega,\vec p))$ are non-analytic at the origin in momentum space, with  
different limits for $\omega \to 0$ at $\vec p= 0$ or for $|\vec p|\to
0$ at $\omega=0$. The first limit yields a plasmon mass, associated
with the damping of plasma oscillations, while the
second is the one that yields the correct finite temperature screening
masses \cite{Das}. Here, in the normal phase the two differ  only for
the sigma meson, by the $n=0 $ contribution proportional to
$\delta_{\omega,0} $ in the equation for $\Pi_\sigma(\omega,T)$, which
can be obtained from the expression for
$\Pi_\sigma(p=(-\imag\omega,\vec p))$ in Ref.~\cite{Brauner:2009gu}
with the additional prescription to set $\omega = 0$ first and then
take $|\vec p | \to 0$.   

The corresponding extra contributions for $\omega = 0$ in the diquark
condensation phase are also given in App.~\ref{sec:rpamass}. 
None of them are really needed here. In particular, the
$\delta_{\omega,0} \Pi^0(T)$ contributions vanish for $T\to 0$,
but they assure that the screening masses extracted from the
propagators agree with the corresponding ones from the effective
potential also at finite temperature, see below.

Setting $\Phi = 1$ and dismissing the temperature
dependent contributions $\delta_{\omega,0} \Pi^0(T)$, the
polarization functions agree with the  
ones from Ref.~\cite{He:2005nk,Xiong:2009zz} for baryon 
chemical potential $\mu_B = 0$ and isospin chemical potential 
$\mu_I = 2\mu$, where $\Pi_\pi$ and $\Pi_\pm $ belong to neutral and
charged pions, respectively.   

The RPA pole masses in the quark-meson-diquark model
in the normal phase are then simply given by the solutions of the
following equations,  
\begin{equation} \label{eq:polnormal}
\begin{array}{rrl}
m_\sigma \, : & \omega^2 =& -m^2 +3\lambda\sigma^2 + \Pi_\sigma(\omega,T) \\[4pt]
m_\pi \, : & \omega^2 =& -m^2 +\lambda\sigma^2 + \Pi_\pi(\omega,T) \\[4pt]
m_\pm \, : & \; (\omega\pm 2\mu)^2 =& -m^2 +\lambda\sigma^2 + \Pi_\pm(\omega,T) \\
\end{array} 
\end{equation}
If we use the mean-field expression in \Eq{eq:MFgcpotential} for the fermionic
pressure with chiral and diquark condensates in the form,
\begin{equation} 
 \Omega_q(T,\mu) = -4T \int\!\! \frac{\d^3q}{(2\pi)^3} \sum_\pm
% \log\left(e^{\frac{E_p^\pm}{T}}+2\Phi+e^{-\frac{E_p^\pm}{T}} \right) \; ,
 \log\left(2 \Big( \cosh \Big(\frac{E_q^\pm}{T}\Big) +\Phi \Big)\right) \, ,
\end{equation}
one immediately verifies that the polarization functions for external
momentum $p=0$, corresponding to the limit $|\vec p|\to 0$ at
$\omega=0$ in the imaginary time formalism, 
\begin{equation}
\label{eqs:Pizero}
\begin{split}
\Pi_\sigma(0,T) &= 2 \frac{\partial}{\partial\sigma^2}
  \Omega_q(T,\mu)\Big|_{d=0}\!+\! 4\sigma^2 \frac{\partial^2}{\partial(\sigma^2)^2}
  \Omega_q(T,\mu)\Big|_{d=0},\\
  \Pi_\pi(0,T) &= 2 \frac{\partial}{\partial\sigma^2}
  \Omega_q(T,\mu)\Big|_{d=0} , \\
  \Pi_+(0,T) &= \Pi_-(0,T)  = 2 \frac{\partial}{\partial d^2}
  \Omega_q(T,\mu)\Big|_{d=0}  . 
\end{split}
\raisetag{3\baselineskip}
\end{equation}
This shows explicitly that the screening masses,
defined by these derivatives of the effective potential, are obtained
as the constant contributions in Eqs.~(\ref{eq:polnormal}) for $\omega =0 $,
\begin{equation} \label{eq:scrnormal}
\begin{array}{rl}
{m_\sigma^\mathrm{sc}}^2  =& -m^2 +3\lambda\sigma^2 + \Pi_\sigma(0,T)\, , \\[4pt]
{m_\pi^\mathrm{sc}}^2  =& -m^2 +\lambda\sigma^2 + \Pi_\pi(0,T)\, , \\[4pt]
{m_\pm^\mathrm{sc}}^2  =& -4\mu^2  -m^2 +\lambda\sigma^2 + \Pi_\pm(0,T) \,
.\\
\end{array} 
\end{equation}
This is true  at all temperatures in the normal phase.  Note also that
because $ \Pi_+(0,T) = \Pi_-(0,T) $,  the baryon chemical potential
$\mu_B=2\mu$ never splits the diquark and antidiquark
screening masses, $m_+^\mathrm{sc}(T,\mu)= m_-^\mathrm{sc}(T,\mu)$.

At any temperature we furthermore verify for $\mu = 0$  that
$\Pi_\pi(\omega,T) = \Pi_\pm(\omega,T) $, {\it i.e.}, pion and
diquark masses are degenerate as they must from $SO(5)$
symmetry. Moreover, the gap equation for the chiral condensate reduces in the 
chiral limit $c\to 0$ to the condition for massless pions, as usual,
and both these observations hold for screening and pole masses, likewise.   

Finally, but maybe most importantly, 
the gap equation for the diquark condensate reads
\begin{equation}
\frac{\partial}{\partial d} \Omega = d \Big( - m^2 + \lambda\sigma^2
-4\mu^2 + 2  \frac{\partial}{\partial d^2}
  \Omega_q(T,\mu) \Big) \stackrel{!}= 0\, ,
\end{equation}
and the critical line $\mu_c(T)$ is defined by the condition that the
terms in brackets vanish for $d=0$ so that a second zero develops
there. This is equivalent to the diquark pole masses being $m_- = 0$ and
$m_+=4\mu$. While their screening masses $m_\pm^\mathrm{sc}$ both
vanish at $\mu = \mu_c$, for the pion and diquark pole masses we have
the general exact zero-temperature relation 
\begin{equation} 
 \Pi_\pm (\omega,0) = \Pi_\pi(\omega\pm 2\mu,0) \;\Rightarrow \quad
 m_\pm = m_\pi \pm 2\mu \, ,  
\end{equation}
in the normal phase where $m_\pi = m_{\pi,0}$ remains independent of $\mu$ until 
$2\mu = m_{\pi,0} $ as required by the Silver Blaze property. 

In contrast, the same relation entails for the degenerate diquark 
screening masses ($\Pi_\pi$ is an even function of $\omega$),
\begin{equation} 
{m_\pm^\mathrm{sc}}^2 =  {m_\pi^\mathrm{sc}}^2 - 4\mu^2 +  
 \Pi_\pi (2\mu,0) - \Pi_\pi(0,0) \, ,  
\end{equation}
which reiterates that diquark and pion screening masses are also
degenerate at $\mu=0$, but that both diquark screening masses 
$m_\pm^\mathrm{sc}$ vanish as $2\mu $ approaches the
($\mu$-independent) pion pole mass $m_\pi $ from below.   

The bottom line is that the onset of diquark condensation at $\mu_B = 2\mu =
2\mu_c(0) $, whatever the screening mass may be,  {\em defines } the
physical zero-temperature pion mass. We will make use of this property
to fix the pion mass in the RG calculation, where the calculation of the
pole mass is more involved.

In the diquark-condensation phase 
%and for $T>0$ one observes a mixing of
the sigma meson mixes with the two diquark modes, {\it
  i.e.}, the respective masses have to be determined from the zeroes
of the determinant of the corresponding $3\times 3$ submatrix in
$\Gamma^{(2)}(\omega)$. 
As in the NJL model \cite{Brauner:2009gu,He:2005nk} one can
verify further exact results from the mass formulas at $T=0$. 
Also in the QMD model at mean-field/RPA level the in-medium pion pole-mass 
is equal to $m_\pi = 2\mu$ above the onset of diquark condensation 
at $2\mu = m_{\pi,0} $. Moreover, one verifies explicitly that 
one of the three modes in the diquark/sigma sector remains exactly
massless in the superfuid phase, also at finite temperature.
This is of course the Goldstone boson corresponding to the
spontaneously broken $U(1)_B$ baryon number.  
Another one becomes degenerate with the pions for large values of the
chemical potential, eventually, reflecting the restoration of chiral
symmetry.  They combine into an $SO(4)$ multiplet
as the chiral condensate vanishes for large chemical potentials. 

\begin{figure}[ht]
\centering
\includegraphics[width=0.48\textwidth]{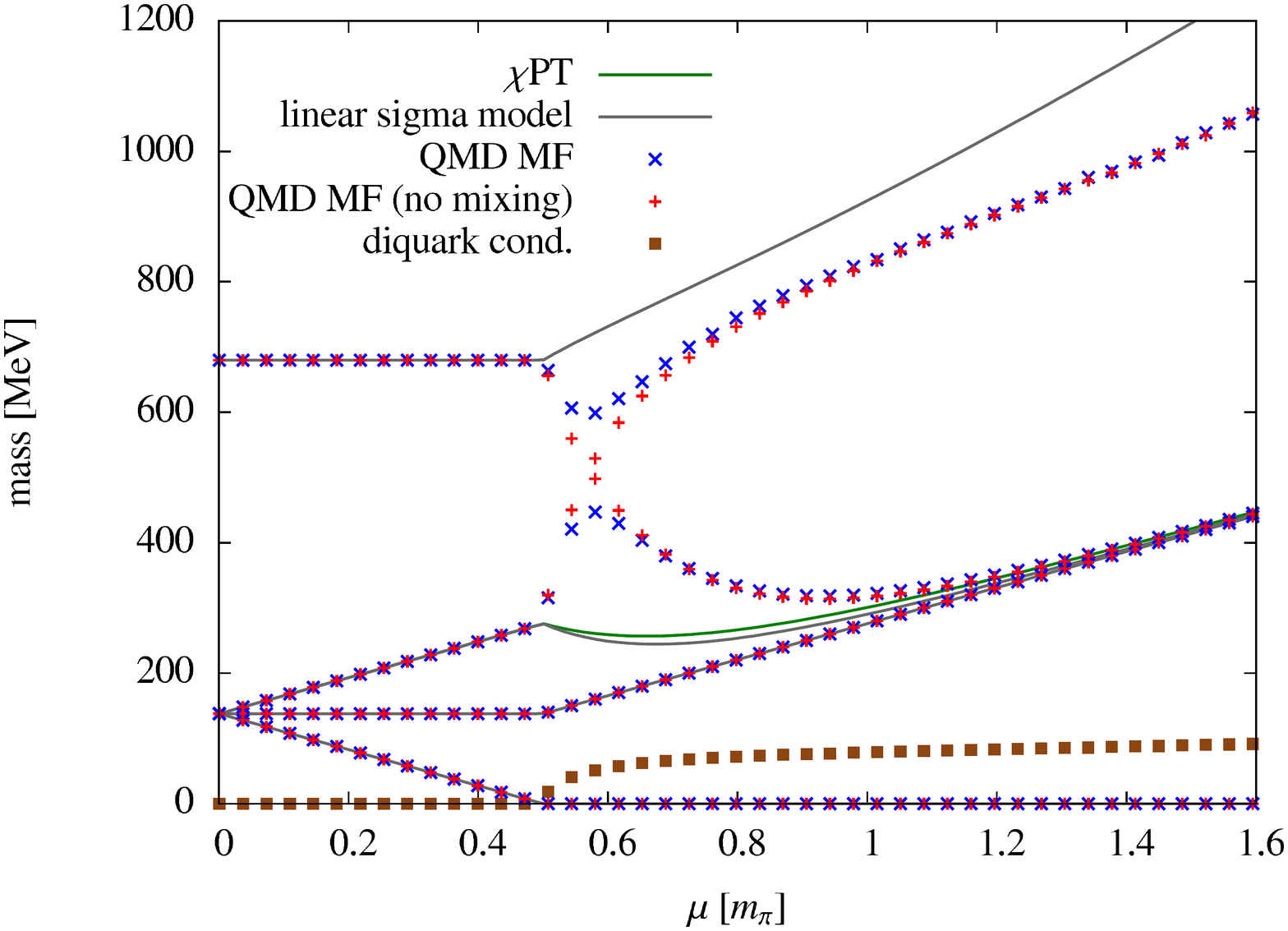}
\vspace{-.2cm}
\caption{Pole-mass spectrum at $T=0$: mean-field/RPA QMD model results
  (with vacuum-term cutoff $\Lambda=600$ MeV),
  for comparison also shown without the effect of diquark/sigma-meson
  mixing in the superfluid phase,  vs.~linear sigma model.  
\vspace{-.5cm} }
\label{fig:mass}
\end{figure}

This is all nicely reflected in the numerical results shown
in Fig.~\ref{fig:mass}. As the RPA pole-mass 
formulas imply, the meson masses stay constant in the normal
phase whereas the diquark masses are split up from the constant
$m_B=m_\pi$ by the terms $\pm\mu_B $ due to their coupling to
the baryon chemical potential  $\mu_B = 2\mu $. 

The diquark and sigma masses in the phase of
diquark condensation show a considerable dependence on the inclusion
of the vacuum term. This can be seen, for example, by comparing the 
QMD model results with vacuum-term cutoff $\Lambda = 600$ MeV 
to those from the linear sigma model, which are identical to the ones
in the no-sea approximation ($\Lambda=0$). 
In the linear sigma model, the pole masses can simply be calculated
from the curvature of the potential. In the normal phase they are 
simply given by the expected constant 
$ m_\pi =m_{\pi,0} $, $m_\sigma=m_{\sigma,0}$, and
$ m_{\pm}=m_{\pi,0}\pm 2\mu $. 
In the phase with diquark condensation ($\mu>\mu_c$), on the other
hand, we obtain for the linear-sigma model masses,
\begin{equation}
\begin{split}
m_\pi=& 2\mu \quad
m_{\Delta_1}=0\\m_{\sigma/\Delta_2}=&\frac{m_{\pi,0}}{\sqrt{2}}\Bigl(-3+y^2+28
x^2 \pm\bigl(3(1-y^2)/x^2\\&+(-3+y^2)^2+40(-3+y^2)x^2+400
x^4\bigr)^\frac{1}{2}\Bigr)^\frac{1}{2} .
\end{split} \label{eq:linsigmass}
\end{equation}
with $x= 2\mu/m_\pi$ and $y = m_\sigma/m_\pi $ as in
\Eq{eq:nonlinsigmamodelcond}.

In addition, Fig.~\ref{fig:mass} 
also shows results of a mass calculation where the mixing terms in the
sigma/diquark sector were neglected. In those results we have explicitly set the
off-diagonal tree-level mixing term $2\lambda\sigma d$ in
\Eq{eq:tlG2sigDelta} to zero, and used  $\Pi_{\sigma\Delta} =  
\Pi_{\sigma\Delta^*} = 0 $ for the polarization functions of
App.~\ref{sec:rpamass} in the
superfluid phase. As illustrated in Fig.~\ref{fig:mass}, 
while the pole masses in the no-sea-approximation are very close to the
$\chi$PT result, the only difference is due to the finite sigma mass
in \Eq{eq:linsigmass},
the more realistic ones with sufficiently large vacuum-term cutoff are
closer to those without any mixing in the crossover region at
intermediate chemical potentials.

%%%%%%%%%%%%%%%%%%%%%%%%%%%%%%%%%%%%%%%%%%%%%%%%%%%%%%%%%%%%%%%%%%%%%%%%%%%%%
%  RG section
%%%%%%%%%%%%%%%%%%%%%%%%%%%%%%%%%%%%%%%%%%%%%%%%%%%%%%%%%%%%%%%%%%%%%%%%%%%%%

\section{FRG Flow equations\label{sec:funRG}}

Quantum and thermal fluctuations are of utmost importance in
particular near phase transitions and are conveniently included within the
framework of the functional renormalization group (FRG) 
\cite{Litim:1998nf,Berges:2000ew, hep-th/0110026, Pawlowski:2005xe, Gies:2006wv,
  Schaefer:2006sr,Braun:2011pp}. In this work we employ a Wilsonian RG
version and investigate the flow equation for the effective average action
pioneered by Wetterich \cite{Wetterich:1992yh}.  The central object in
this approach is the renormalization scale $k$ dependent effective
average action $\Gamma_k[\Phi]$, where $\Phi$ generically represents the 
set of all quantum fields of the theory.  The effective average action
interpolates between the microscopic classical action at some ultraviolet 
(UV) cutoff scale $k=\Lambda$, at which fluctuations of essentially all  
momentum modes are suppressed, and the effective action of the  
full quantum theory in the infrared (IR), for $k\to0$, which then
includes all quantum and thermal fluctuations.  The scale-dependence
is described by the Wetterich flow equation,  
\begin{equation}
\label{eq:Wettericheq}
\partial_t \Gamma_k\equiv k\partial_k
\Gamma_k[\Phi]=\frac{1}{2}\text{Tr}\left\{\,\partial_t R_k
  (\Gamma^{(2)}_k+R_k)^{-1}\right\}\ ,
\end{equation}
which involves a momentum- and scale-dependent regulator $R_k$, whose 
precise form is not fixed but leaves a considerable flexibility. The
role of the regulator $R_k$ is to suppress the fluctuations of modes
with momenta below the renormalization scale $k$, and the flow
equation is UV as well as IR finite.  $\Gamma^{(2)}_k[\Phi]$ 
are the second functional derivatives of the effective average action
with respect to all the fields at scale $k$.  The functional trace
represents a one-loop integration typically evaluated in momentum
space and includes the sum over all fields and their internal and
space-time indices as well, with standard modifications for fermionic fields.  
It contains the full field and $k$-dependent
propagators of the regulated theory with cutoff $R_k$, the inverse of  
$\Gamma^{(2)}_k[\Phi ] + R_k$.  
In order to solve the flow equation an initial microscopic action
$S=\Gamma_{k=\Lambda}$ at some UV scale $\Lambda$ has to be specified.
Truncating the effective action to a specific form,  
the functional equation can be converted into a closed set of
(integro-)differential equations, but will in general also introduce
some regulator dependence in the flow. The choice of an
optimized regulator minimizes this regulator dependence for physical
observables. As bosonic (fermionic) regulators $R_{k,B}$ ($R_{k,F}$) we
choose 

\vspace{-.1cm}

\begin{equation}
\label{eq:regulators}
\begin{split}
R_{k,B}(\vec p)&=(k^2-\vec p^2)\theta(k^2-\vec p^2),\\
R_{k,F}(\vec p)&=-\imag \vec{p}\cdot\vec\gamma
\left(\sqrt{\frac{k^2}{\vec p^2}} -1\right)\theta(k^2-\vec p^2),
\end{split}
\end{equation} 
which are three-momentum analogues of the optimized Litim regulators
\cite{Litim:2001up}. With this choice the three-momentum integration
becomes trivial and the remaining Matsubara sums can be evaluated
analytically. Furthermore, this choice leaves the semilocal
$U(1)$-symmetry of the Lagrangian unaffected, analogous to
\cite{Diehl:2009ma}, where the chemical potential acts like the
zero-component of an Abelian gauge field. In addition, these
regulators have their precise counterparts in specific three-momentum
regulators for proper-time flows which then lead to identical 
flow equations, c.f. Appendix~\ref{sec:ptRG}. On the other hand,
especially in a relativistic system 
the fact that the zero-component of the momentum is not regulated can
potentially be problematic. %%Ref here for example??

%\vspace{-.1cm}

% the local potential approximation (LPA) .
Our ansatz for the effective average action in leading-order
derivative expansion, where all wave-function renormalization factors
are neglected and only the scale-dependent effective potential $U_k$
is taken into account, reads
\begin{equation}
\label{eq:ansatzgammak}
\Gamma_k=\left.\int d^4x \mathcal{L}_\text{PQMD}\right|_{V+c\sigma
  \to U_k}\ .
\end{equation}
This means that we use $\mathcal{L}_\text{PQMD}$ from
\Eq{eq:LagrangianPQM}, but replace the
meson/diquark potential $V(\vec\phi)$ of the $O(6)$ linear sigma model
from \Eq{eq:Vlinsig} therein by $U_k-c\sigma$. The explicit symmetry
breaking term $-c\sigma $ does not affect the flow and is thus not
part of $U_k$ but added after the RG evolution to the full effective
potential again. At $\mu=0$, the scale-dependent $U_k$ then only depends on
the modulus of $\vec\phi=(\sigma,\vec{\pi},\text{Re}\,\Delta,\text{Im}\,\Delta)^T$. At  
non-vanishing chemical potential, however, we only have $SO(4)\times
SO(2)$ symmetry and must therefore allow it to depend on two
invariants, {\it i.e.}, $U_k\equiv U_k(\rho^2,d^2)$ where $\rho^2 =
\sigma^2+\vec\pi^2$, and $d^2 = |\Delta|^2$ as before. For $\mu\to 0$
we recover $SO(6)$ invariance, of course, so that $U_k$ then 
depends only on the combination $\phi^2 = \rho^2 + d^2$ again.

With the constant field configurations $\sigma=\rho$, $\vec\pi=\vec 0$, 
$\text{Re}\, \Delta =  d$, $\text{Im}\,\Delta=0$ we obtain for the
bosonic second functional derivative of the effective action
%To obtain the bosonic contributions  flow we consider a constant
%field configuration with $\vec\pi=\vec 0$, $\sigma=\rho$,
%$\text{Re}\,\Delta=d$, $\text{Im}\,\Delta=0$ and find for
%$\frac{\delta^2\Gamma_k}{\delta \phi_i(p)\delta
 % \phi_j(p')}=(\Gamma^{(2)}_B)_{ij}\delta(p+p')$, where
\begin{widetext}
\begin{equation}%\mathbbm{1}_{3\times 3}
\label{eq:gamma2b}
\Gamma^{(2)}_{k,B}\!=\!\left(\begin{array}{cccccc}
p^2+2\Ur &0 &0 &0 &0&0\\
0&   p^2+2\Ur &0 &0 &0 &0\\
0& 0&  p^2+2\Ur &0 &0 &0\\
0& 0& 0& p^2+2 \Ur+4 \rho^2 \Urr&4 \rho d \Urd&0\\
0& 0& 0& 4 \rho d \Urd&p^2+2\Ud+4 d^2 \Udd-4\mu^2& -4 \mu p_0\\
0& 0& 0& 0&4 \mu p_0&p^2+2\Ud-4 \mu^2
\end{array}\right)\! ,
\end{equation}
\end{widetext}
where we have introduced short-hand notations for the
derivatives  of the potential with respect to the fields defined as
 $\Ud \equiv \partial U_k/\partial d^2$, $\Ur \equiv
\partial U_k/\partial \rho^2$ and later we will also use $\Up \equiv
\partial U_k/\partial \phi^2$. Higher order derivatives are labeled
with higher order indices accordingly, e.g., $\Urd \equiv \partial^2
U_k/\partial \rho^2\partial d^2$.  The alert reader will have  noticed that
\Eq{eq:gamma2b}
 agrees with Eqs.~(\ref{eq:tlG2pions}) and (\ref{eq:tlG2sigDelta})
upon working out these derivatives, if we replace $U_k$ back to
$\lambda (\phi^2-v^2)^2/4 $ which is what we use at $k=\Lambda$. 

In the fermionic sector we find analogously,
\begin{equation}
  \Gamma^{(2)}_{k,F}=\left(\begin{smallmatrix}-\imag \slashed p-\imag a_0 \gamma^0 + g \rho -\gamma^0 \mu &g\gamma^5 d\\
      -g\gamma^5 d &-\imag \slashed p -\imag a_0 \gamma^0+ g\rho
      +\gamma^0\mu\end{smallmatrix}\right)\otimes
  \mathbbm{1}_{2\times2}\ .
\end{equation}

\noindent
To these expressions we add the respective regulators in
\Eq{eq:regulators} before they are being inverted and inserted into 
the Wetterich equation \Eq{eq:Wettericheq}, in order to obtain the
flow equation for the effective potential. 
Replacing the zero-components $p_0$ of the momenta 
by periodic (antiperiodic) Matsubara frequencies $\omega_n= 2\pi
nT$ ($\nu_n= (2 n+ 1) \pi T$), and upon performing the spatial
momentum integrations, the corresponding bosonic and fermionic
contributions to the flow for the effective potential are
then given by the following Matsubara sums,
%%%%%%%%%%%%%%%%%%%%%%%%%%%%%%%%%%%%%%%%%%%%%%%%%%%%%%%%%%%%%%%%%%%%%%%%%%%%%
%original
% \begin{widetext}
% \begin{align}
%   \partial_t U_{k,B}&=\frac{1}{2\mathcal{V}}\Tr\,\left\{ \partial_t R_{k,B} ({ \Gamma_{k,B}^{(2)}}+R_{k,B})^{-1}\right\}=\frac{k^5 T}{6\pi^2}\sum_n\biggl(\frac{3}{{p^0}^2+k^2+2U_{r}}+\Tr\, A^{-1}(\vec p^2\to k^2)\biggr)\notag\\
%   &=\frac{k^5 T}{6 \pi^2}\sum_n\biggl(\frac{3}{{p^0}^2+k^2+2U_{,r}}+\frac{\alpha_4 {p^0}^4+\alpha_2 {p^0}^2+\alpha_0}{\beta_6 {p^0}^6+\beta_4 {p^0}^4+\beta_2 {p^0}^2+\beta_0}\biggr)\label{eq:bosonicflow}\\
%   \partial_t
%   U_{k,F}&=-\frac{1}{\mathcal{V}}\text{Tr}\,\left\{\partial_t
%     R_k(\Gamma^{(2)}_{k,F}+R_{k,F})^{-1}\right\}=-\frac{8k^5
%     T}{3\pi^2}\sum_n
%   \frac{d^2g^2+k^2-\mu^2+{(p^0+\theta)}^2+g^2\rho^2}
%   {({(p^0+\theta)}^2+\gamma_+^2) ({(p^0+\theta)}^2+\gamma_-^2)}\label{eq:fermionicflow}, 
% \end{align}
% \end{widetext}
%%%%%%%%%%%%%%%%%%%%%%%%%%%%%%%%%%%%%%%%%%%%%%%%%%%%%%%%%%%%%%%%%%%%%%%%%%%%%
\begin{widetext}
\begin{align}
  \partial_t U_{k,B} =  & \frac{k^5 T}{6 \pi^2}\sum_{n\in Z}\biggl( \frac{3}{\omega_n^2+k^2+2\Ur}
+ \frac{\alpha_2 (\omega_n^2)^2+\alpha_1 \omega_n^2+\alpha_0}
{(\omega_n^2)^3+\beta_2 (\omega_n^2)^2+\beta_1 \omega_n^2+\beta_0} 
\biggr) \ ,
\label{eq:bosonicflow}\\
  \partial_t
  U_{k,F} =&-\frac{8k^5
    T}{3\pi^2}\sum_{n \in Z}
  \frac{(\nu_n+a_0)^2 + k^2 + g^2 \phi^2  -\mu^2 }
  { \big(  (\nu_n+a_0)^2+{E^+_k}^2 \, \big)  \big(
      (\nu_n+a_0)^2+{E^-_k}^2 \, \big) } \ ,
\label{eq:fermionicflow} 
\end{align}
\end{widetext}
where we have introduced % ${E}^\pm_k  =\sqrt{g^2 d^2 + {\epsilon^\pm_k}^2}$,
%$\epsilon^\pm_k = \epsilon_k \pm \mu $ 
${E}^\pm_k  =\sqrt{g^2 d^2 + (\epsilon_k \pm \mu)^2}$,
and $\epsilon_k =\sqrt{k^2+ g^2\rho^2}$ analogous to the notations of
Sec.~\ref{sec:MF}.  
The numerator of the second term in the bosonic flow,
\Eq{eq:bosonicflow}, is a quadratic polynomial in $\omega_n^2$ with
three coefficient functions $\alpha_{i}$, while the
denominator is a cubic polynomial in standard form, with coefficient
functions $\beta_{i}$ and leading coefficient $\beta_3 = 1$.
These coefficient functions $\alpha_{i}$ and  $\beta_{i}$ depend on
renormalization scale, chemical potential, fields and the derivatives
of the potential. They are obtained straightforwardly from the trace
of the inverse of the $3 \times 3$ submatrix of the bosonic 2-point
function in \Eq{eq:gamma2b} corresponding to the sigma and diquark
directions in field space, and they are listed explicitly in Appendix~
\ref{sec:bosonicflowcoefficients} for completeness. With the roots of
the denominator which we denote as  $\omega_{n,0}^2=-z_i^2$,
$i=1,...,3$, we can evaluate all Matsubara sums analytically by virtue
of the residue theorem in a standard way. 

Hence, the final flow equation for the effective potential of the PQMD
model is the sum of the bosonic and fermionic flow and reads
explicitly,
\begin{widetext}
\begin{equation}
\label{eq:fullflowfinal}
\begin{split}
  \partial_t U_k=
  \frac{k^5}{12\pi^2}&\left\{\frac{3}{E^\pi_k}\coth\left(\frac{E^\pi_k}{2T}\right)
    +\sum_{i=1}^3  \frac{\alpha_2 z_i^4-\alpha_1
   z_i^2+\alpha_0} {(z_{i+1}^2-z_i^2) (z_{i+2}^2-z_i^2)}
    \frac{1}{z_i} \coth\left(\frac{z_i}{2T}\right)\right.\\
  &\hspace{3cm} \left.-\sum_{\pm}\frac{8}{E^\pm_k}\left(1 \pm
      \frac{\mu} {\sqrt{k^2+g^2\rho^2}}\right) 
   \, \Big( 1 - 2 N_q(E_k^\pm;T,\Phi) \Big) 
%\frac{\sinh(\beta E_k^\pm)}{\cos(\beta a_0)+\cosh(\beta E_k^\pm)}
\right\}\ , 
\end{split}
\end{equation}
\end{widetext}
where $E^\pi_k=\sqrt{k^2+2 \Ur}\,$, and $N_q(E;T,\Phi)$ are  
the Polyakov loop enhanced quark occupation numbers from
\Eq{eq:PLenhancedoccnumbers}. Without a diquark  
condensate, {\it i.e.}~by setting explicitly $\Delta=0$, 
we can write down an $SO(6)$-symmetric flow equation for $U_k(\phi )$,
if we set $\Up\equiv \Ur= \Ud$. \Eq{eq:fullflowfinal} then reduces
to the more familiar looking form,
\begin{widetext}
\begin{equation}
\label{eq:floweqdelta0}
\begin{split}
  \partial_t
  U_k=\frac{k^5}{12\pi^2}&
  \left\{\frac{3}{E_k^{\pi}}\coth\left(\frac{E_k^{\pi}}{2T}\right)
+\frac{1}{E_k^\sigma}\coth\left(\frac{E_k^\sigma}{2T}\right)
    +\frac{1}{E_k^{\pi}}\coth\left(\frac{E_k^{\pi}-2\mu}{2T}\right)
    +\frac{1}{E_k^{\pi}}\coth\left(\frac{E_k^{\pi}+2\mu}{2T}\right) \right.\\
%  &\left. -\frac{4N_c N_f}{\epsilon_k} \,
  &\hspace{.5cm} \left. -\frac{16}{\epsilon_k} \,
  \Big(\left. 1-N_{q}\left(\epsilon_k-\mu;T,\Phi\right) 
    -N_{q}(\epsilon_k+\mu;T,\Phi)\right. \Big)\right\}, 
\end{split}
\end{equation}
\end{widetext}
with single-particle energies for mesons/diquarks
$E_k^{\pi}=\sqrt{k^2+2\Up}$ and sigma $E_k^\sigma=\sqrt{k^2+2 \Up  
  +4\phi^2 \Upp}$.
% and Polyakov loop enhanced quark/antiquark occupation numbers
%$N_q$ defined in Eq. (\ref{eq:PLenhancedoccnumbers}).
Except for the change in the number of active degrees of freedom  
contributing to this flow, and the isospin-like chemical potential
coupling to one pseudo-Goldstone boson pair, the $SO(6)$ symmetric flow
equation here is entirely analogous the one of the PQM model in the
three-color case, see {\it e.g.}, \cite{Herbst:2010rf,Skokov:2010uh,
  Skokov:2010wb}.  
For the three-color PQM model with isospin chemical potential one must
allow for pion condensation, however, and
then arrives at a flow equation \cite{KazuhikoEtAlInPrep}
analogous to our \Eq{eq:fullflowfinal}.

In the following sections we present numerical solutions to the
flow equation (\ref{eq:fullflowfinal}).  The full effective potential
depends in general on three condensates which hampers its numerical
solution enormously. In order to proceed we restrict ourselves in this
work to the two-color QMD model and neglect the influence of the
Polyakov-loop by setting $\Phi=1$ in the flow and postpone the full
PQMD model solution for a later analysis. For the first time we
generalize the one-dimensional grid solution technique to two
dimensions. Details of the numerical procedure and the parameter
fixing can be found in Appendix~\ref{sec:paramfixing}.

%%%%%%%%%%%%%%%%%%%%%%%%%%%%%%%%%%%%%%%%%%%%%%%%%%%%%%%%%%%%%%%%%%%%%%%%%%%%%

\begin{figure}[t]
\centering
\includegraphics[width=0.48\textwidth]{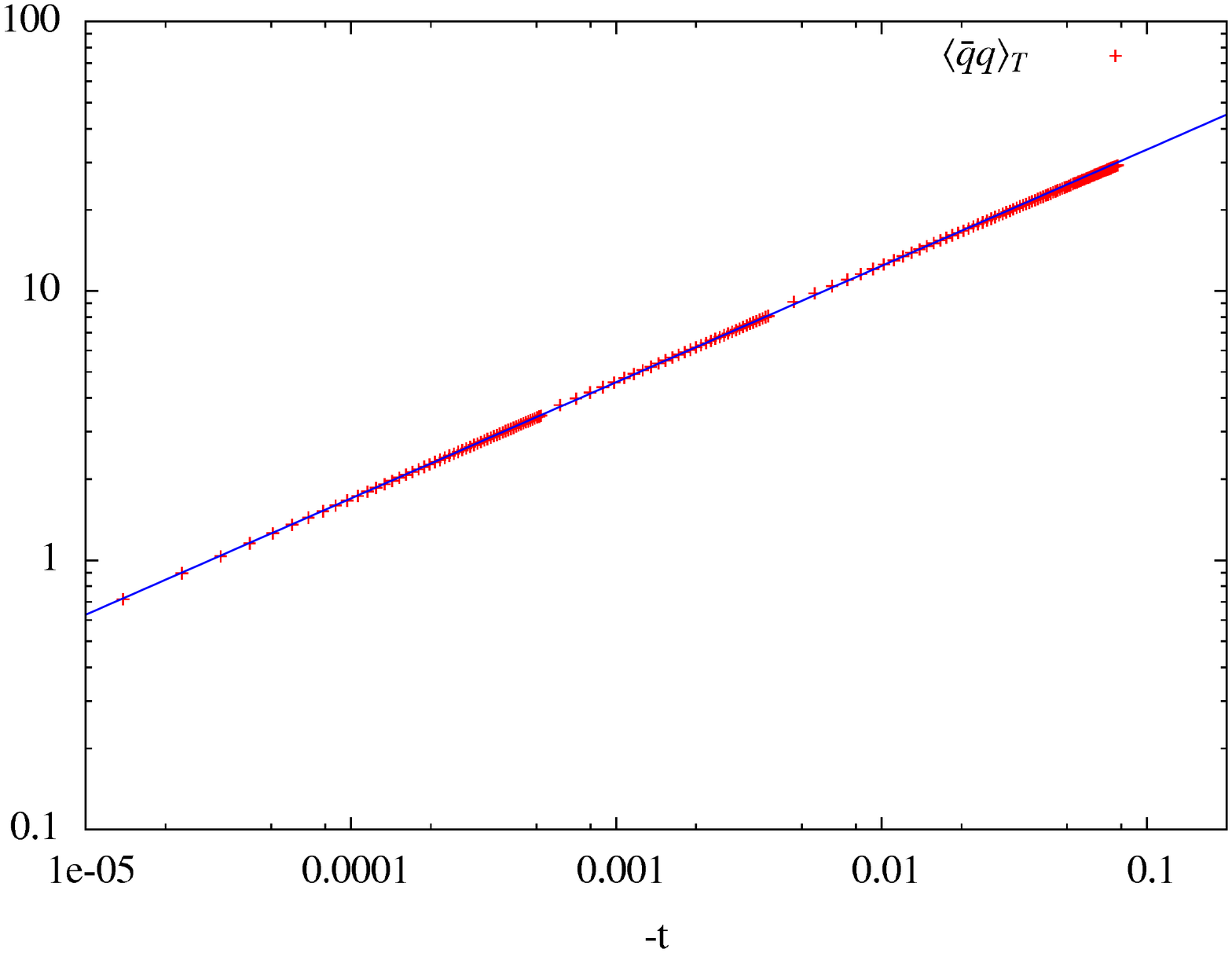}
\vspace{-0.2cm}
\caption{Fit to chiral condensate for critical exponent $\beta$.
\vspace{-0.2cm}}
\label{fig:expbeta}
\end{figure}

\subsection{Critical exponents $\beta$ and $\delta$}
\label{sec:O6scale}

Without diquark condensation for vanishing chemical potential, and
Polyakov-loop variable $\Phi=1$, the $SO(6)$ symmetric flow in
\Eq{eq:floweqdelta0} further simplifies to
\begin{equation}
\label{eq:flowmu0}
\begin{split}
\partial_t U_k= \frac{k^5}{12 \pi^2} & \left\{\frac{5}{E_k^\pi}
\coth\left(\frac{E_k^\pi}{2 T}\right) +\frac{1}{E_k^\sigma} \coth\left(
  \frac{E_k^\sigma}{2 T}\right) \right. \\ 
& \hspace{1cm} \left. %-\frac{4 N_c  N_f}{\epsilon_k }
-\frac{16}{\epsilon_k }
\tanh\left(\frac{\epsilon_k}{2 T} \right) \right\}\ .
\end{split}
\end{equation}

At $\mu=0$ the diquarks are degenerate with the pions which leaves us with
the $N_c=2$ analogue of the familiar three-color QM model flow
equation \cite{Braun:2003ii,Schaefer:2004en} except that there are now
five pseudo-Goldstone bosons instead of the usual three pions. 
%In addition,  
%for $\mu=0$ the flow \Eq{eq:flowmu0} coincides with
%the full flow \Eq{eq:floweqdelta0} when $d$ is set to zero.

\begin{figure}[t]
\centering
\includegraphics[width=0.48\textwidth]{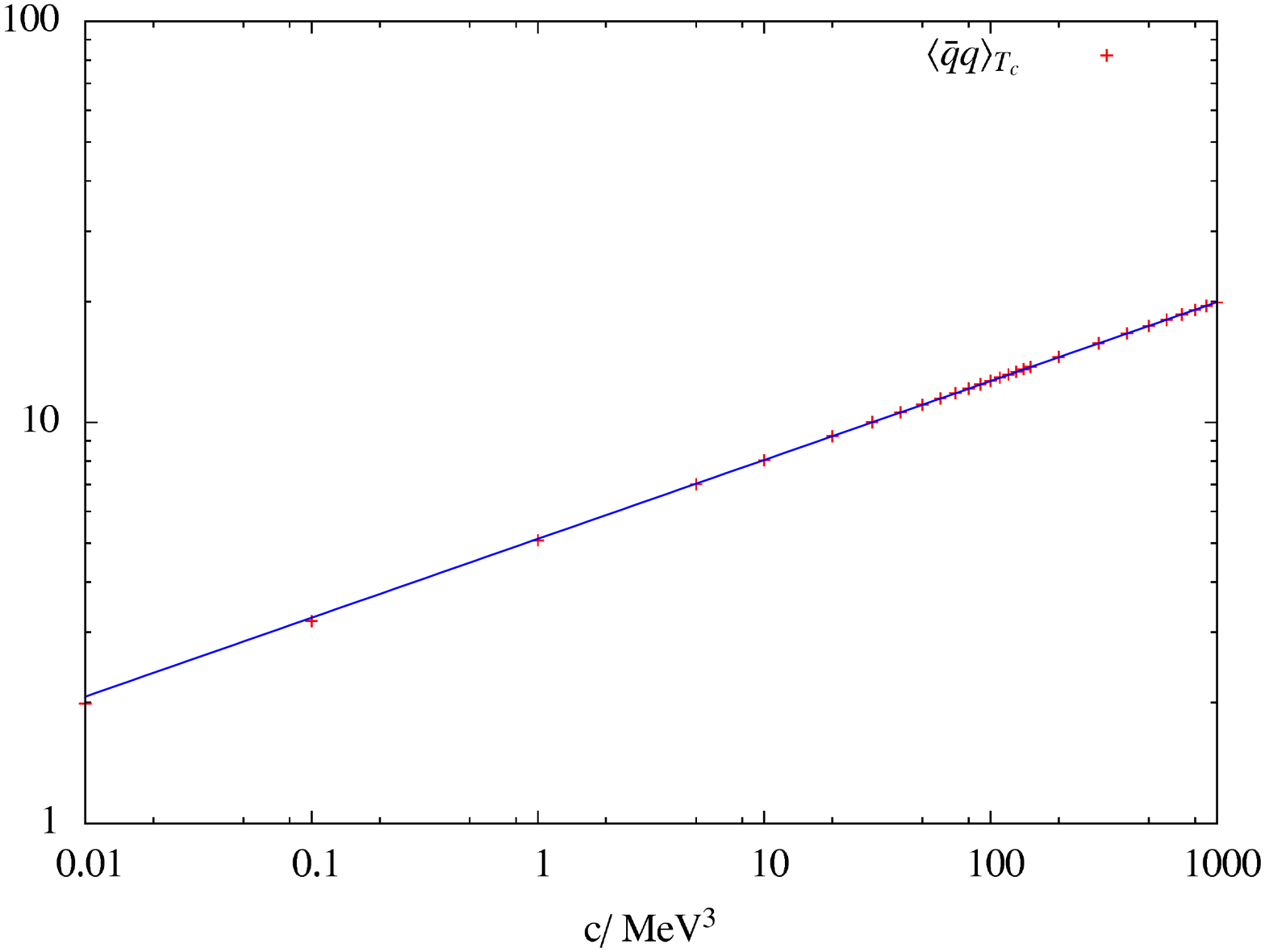}
\vspace{-0.2cm}
\caption{Fit to chiral condensate for  
 critical exponent $\delta$.
\vspace{-0.2cm}}
\label{fig:expdelta}
\end{figure}

The study of $O(4)$ universality and scaling in the three-color QM model
has a long history by now
\cite{Berges:1997eu,Schaefer:1999em,Bohr:2000gp,Stokic:2009uv,Braun:2010vd}. 
Here we can analogously check the symmetry breaking patterns discussed
in Section \ref{sec:chsymbreaking} by computing the corresponding 
critical exponents. 
As discussed there, for $\mu=m=0$, the $SU(4)\simeq SO(6)$ dynamically
breaks down to $Sp(2) \simeq SO(5)$ so that we expect a finite
temperature phase transition in the three-dimensional $O(6)$
universality class. The critical exponent $\beta$ can be extracted
from the dependence of the chiral condensate on the reduced
temperature $t=(T-T_c)/T_c$ in the chiral limit, whereas the exponent
$\delta$ governs the dependence of the chiral condensate at $T_c$ on
the quark mass $m_q$ or correspondingly on the explicit
symmetry-breaking parameter $c$, 
\begin{equation}
\langle \bar q q\rangle_T\sim(-t)^\beta , \quad \langle \bar q
q\rangle_{T_c}\sim c^{1/\delta} .
\end{equation}
With the usual %scaling and hyperscaling relations 
two-exponent scaling all other critical
exponents are then obtained from these two.
Here we find critical exponents $\beta=0.4318(4)$ and $\delta=5.08(8)$
from the solution of the 1$d$ flow equation (\ref{eq:flowmu0}) via the
Taylor expansion method. The given errors are statistical errors
extracted from the fit. The corresponding fits are shown in
Figs.~\ref{fig:expbeta} and \ref{fig:expdelta}, respectively. 
Literature values for these exponents obtained from Monte-Carlo
simulations are given by $\beta = 0.425(2)$ and $\delta = 4.77(2)$
\cite{Holtmann:2003he}. At the leading order derivative expansion
employed here, we should not expect to reproduce these values, however.
The more appropriate benchmark here should be the functional 
renormalization group result for the $O(6)$ model in leading 
order derivative expansion \cite{Litim:2002cf}. 
In absence of wave-function renormalizations there is no 
anomalous dimension for the fields and their critical exponent
therefore vanishes, $\eta = 0$. Then the  hyperscaling relations,  
\begin{equation}
\delta = \frac{d+2 -\eta}{d-2+\eta} \, , \quad \beta = \frac{\nu}{2}
\big( d-2+\eta \big) \, ,
\end{equation}
immediately entail that $\delta = 5$ and $\beta = \nu/2$ in $d=3$
dimensions. With the correlation-length critical exponent 
$\nu = 0.863076 $ from Ref.~\cite{Litim:2002cf} this corresponds to
$\beta = 0.4315 $, and both our values are in agreement 
with these two within our errors.

%%%%%%%%%%%%%%%%%%%%%%%%%%%%%%%%%%%%%%%%%%%%%%%%%%%%%%%%%%%%%%%%%%%%%%%%%%%%%
\subsection{Phase diagram without diquark fluctuations}
\label{sec:pwwodq}

\begin{figure}[hb]
\centering
\includegraphics[width=0.48\textwidth]{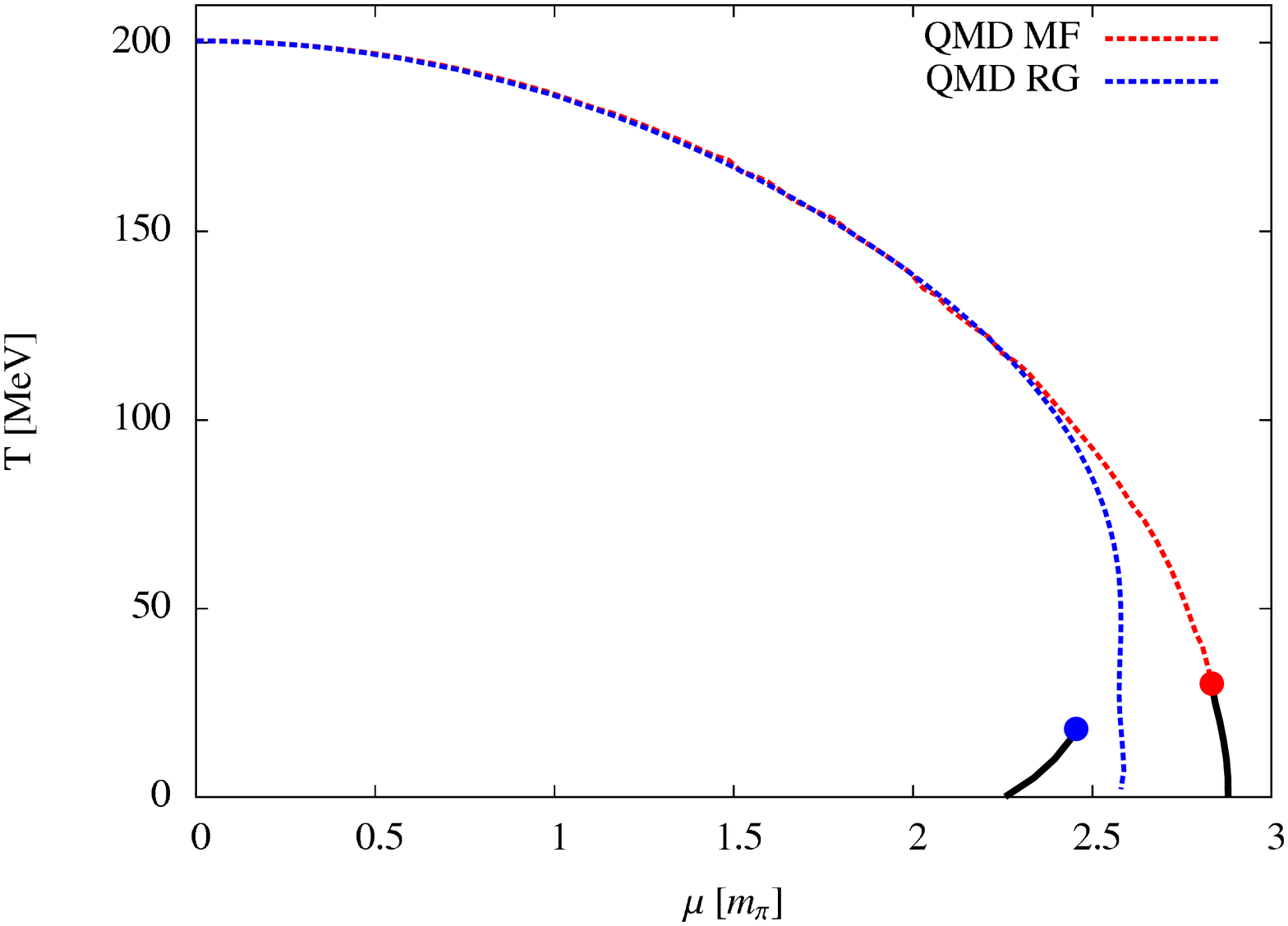}

\vspace{-.4cm}
\caption{QMD model phase diagram without baryonic degrees of freedom
  ($d=0$): 1$d$ RG result vs. MF ($\Lambda=600$ MeV).
\vspace{-.2cm} }
\label{fig:pdd0}
\end{figure}

Before we discuss the solutions to the full flow equation
(\ref{eq:fullflowfinal})  for the effective potential with
fluctuations of both condensates included, it might be instructive to
illustrate the influence of fluctuations on the standard
quark-meson-model-like phase diagram without baryonic degrees of
freedom. The phase diagram obtained from our solutions to the
$SO(6)$-symmetric flow equation (\ref{eq:floweqdelta0}) in the
($T,\mu$)-plane is compared to the mean-field results from
Sec.~\ref{sec:MFvac} in Fig.~\ref{fig:pdd0}. The mean-field
solutions there were obtained from \Eq{eq:MFgcpotential} with $d^2 =
|\Delta|^2=0$ and with a vacuum term cut off at $\Lambda=600$ MeV which
is sufficiently large for a reasonable comparison, see
Sec.~\ref{sec:MFvac}. The parameters were chosen so as to match the
$\mu = 0$ chiral transition temperatures (rather than the 
sigma mass) in addition to pion decay constant and pion mass, as explained in
Appendix~\ref{sec:paramfixing}.

Again, the resulting phase diagram with fluctuations shows
the typical form of the QM model for  $N_c=3$. It has a
critical endpoint at $\mu\approx 2.5 m_\pi$ as compared to that at 
$\mu\approx 2.8 m_\pi$ in the mean-field calculation. The dotted
chiral-crossover lines are again simply the half-value curves of the
chiral condensate. Except for the shift of the critical endpoint and
the crossover-line in the range of chemical potentials between $2.5$
and $2.8$ $m_\pi$, the chiral-crossover line from the RG solution
closely follows the mean-field result.  As discussed in
Sec.~\ref{sec:DiqCond} already, they are both equally wrong for baryon
chemical potentials of the order of the baryon mass and above, as we
have neglected the essential dynamics of the baryonic diquarks in the
superfluid phase.
 
The advantage here as compared to the three-color case is
that it is much more straightforward to include baryonic degrees of
freedom, by properly including the diquarks, both at mean-field level
and in the full flow equation (\ref{eq:fullflowfinal}). All we have to
do is to solve this equation in the higher-dimensional space of
invariants in the fields representing both possible order parameters
and their fluctuations at the same time.

%%%%%%%%%%%%%%%%%%%%%%%%%%%%%%%%%%%%%%%%%%%%%%%%%%%%%%%%%%%%%%%%%%%%%%%%%%%%%
\subsection{FRG pole mass and flow of the 2-point function\label{sec:flow2ptfn}}

The fact that the Silver Blaze property links the onset of diquark condensation
to the zero-temperature pion mass represents a strong constraint which
has to hold in the (P)QMD model with fluctuations also. As we have already
illustrated at mean-field level in Section \ref{sec:polemass}, a proper
definition of meson and baryon masses is absolutely crucial for this
exact property of the theory. In particular, we have seen explicitly that
the behavior of widely used screening masses is unphysical in this
regard. In the mean-field calculations, the difference between the
pion's pole and screening masses at zero temperature and with identical
model parameters can be as much as 30\%, for example. Adjusting the
parameters to the more physical pole mass instead of the common
procedure in these models has a considerable influence on the results.
So this is more than an academic exercise.

Therefore, we propose a simple procedure to obtain pole masses
suitable for the FRG framework: 
As an extension to the flow equation for the effective
potential in the leading order derivative expansion, 
%or local potential approximation (LPA), 
we solve the flow equations 
for the 2-point functions of mesons and diquarks
using the field and scale dependent but momentum independent 3- and 4-point
vertices obtained from the effective potential.  This ensures that the
flow equations for the 2-point functions at zero momentum reduce to
those for the mass terms in the effective potential and that the
screening masses obtained from the flows of 2-point functions and
effective potential are therefore the same by construction. This
truncation for the flow of the 2-point functions most naturally
extends that of the effective potential, and thus provides a simple
alternative to other approaches such as the BMW approximation
\cite{Benitez:2011xx} where the same relation with the effective potential
typically arises as an additional requirement.  

In this section we outline the derivation and solution methods for
the flow equation of the pion 2-point function $\Gamma^{(2)}_{\pi\pi}$
at  $T=\mu=0$ as an example which will allows us to define a pion pole
mass in the vacuum. %For simplicity we restrict ourselves to here. 
We consider $N_f$ flavors of quark with $N_c$ colors coupled to an
$O(N)$-symmetric bosonic sector for combinations of $N_f$, $N_c$ and
$N$ where this is possible.  

The flow equation for the (field dependent) 2-point function is given
by the second functional derivative of the original flow equation 
(\ref{eq:Wettericheq}) which in our case is,
\begin{widetext}
\begin{equation}
\label{eq:flow2ptfn}
\begin{split}
\partial_t \Gamma^{(0,2)}_{ij}(p;\phi)=\int \frac{\d^4
  q}{(2\pi)^4}&  \partial_t R^B(q)_{kl}\Bigl( 
G^B(q)_{lm}\Gamma^{(0,3)}_{mnj}
G^B(q-p)_{nr}\Gamma^{(0,3)}_{rsi}G^B(q)_{sk}-\frac{1}{2} G^B(q)_{lm}
\Gamma^{(0,4)}_{mnij}G^B(q)_{nk}\Bigr)\\ 
&-2 \, \mbox{tr} \, \partial_t R^F(q)\Bigl( G^F(q)\Gamma^{(2,1)}_{j}
G^F(q-p)\Gamma^{(2,1)}_{i}G^F(q)-\frac{1}{2} G^F(q)
\Gamma^{(2,2)}_{ij}G^F(q)\Bigr)\, ,  
\end{split}
\end{equation}
\end{widetext}
where $G^{B}(p)=\big(\Gamma^{(0,2)}_k(p;\phi)+R^B_k(p)\big)^{-1}$ and
$G^{F}(p)=\big(\Gamma^{(2,0)}_k(p;\phi)+R^F_k(p)\big)^{-1}$. To
solve the flow equation (\ref{eq:flow2ptfn}) one needs $3^\mathrm{rd}$
and $4^\mathrm{th}$ derivatives of $\Gamma_k[\phi]$ which we denote
generically by
$\Gamma_k^{(n,m)}$ where the first(second) superscript counts
the number of fermionic(bosonic) derivatives. In order for the
limit $p\to 0$ to be consistent 
with the truncation used for the flow equation of the effective
potential we obtain those higher $n$-point vertex functions from
the same scale-dependent effective action. In leading order derivative
expansion, Eqs.~(\ref{eq:ansatzgammak}), (\ref{eq:fullflowfinal}) or
(\ref{eq:flowmu0}) for $\mu=0$, the 3- and 4-point functions are then
momentum independent, and the only dependence on the external momentum
comes from the propagators themselves.  
For convenience we choose coordinates $\phi_i=\phi
\delta_{i 1}$, {\it i.e.}, $\sigma = \phi$ and the others zero, so that
one has explicitly for the quark-boson vertices with constant Yukawa
couplings,
\begin{equation}
\Gamma^{(2,1)}_0=g\, ,\quad\Gamma^{(2,1)}_{j\neq 0}=\imag g \gamma^5
\tau_j\, , \quad \Gamma^{(2,2)}_{ij}=0\ .
\end{equation}
The three and four-boson vertices are extracted from the respective 
derivatives of the $k$-dependent effective potential $U_k$. Here this is a
function of $\phi^2$ and we simply write $U_k''(\phi^2) \equiv 
\Upp $ etc. instead of our index notation above, 
\begin{equation}
\begin{split}
\Gamma^{(0,3)}_{ijm}&=4 \phi\,  U_k'' \, (\delta_{ij}\delta_{m
  1}+\delta_{jm}\delta_{i 1}+\delta_{im}\delta_{j 1})\\
                   &\quad +8\phi^3\,  U_k^{(3)} \, \delta_{i
  1}\delta_{j 1}\delta_{m 1} \, , 
\end{split}
\end{equation}
\begin{equation}
\begin{split}
\Gamma^{(0,4)}_{ijmn}&=4 \, U_k'' \,
(\delta_{ij}\delta_{mn}+\delta_{in}\delta_{jm}+\delta_{jn}\delta_{im})\\
    &\quad + 8 \phi^2\,  U_k^{(3)}\, (\delta_{ij}\delta_{m 1}\delta_{n 1}
          +\delta_{jm}\delta_{i 1}\delta_{m 1}+\delta_{mn}\delta_{i
  1}\delta_{j 1}\\ 
 &\hspace{2cm} +\delta_{jn}\delta_{i 1}\delta_{m
  1}+\delta_{in}\delta_{j 1}\delta_{m 1})\\
&\quad+16 \phi^4 \, U_k^{(4)}\, \delta_{i 1}\delta_{j 1}\delta_{m
  1}\delta_{n 1} \, .
\end{split}
\raisetag{\baselineskip}
\end{equation}
For the calculation of the boson masses we use their rest frame,
setting the spatial external momentum $\vec p=0$ in
\Eq{eq:flow2ptfn}. In this frame the spatial momentum integrals with
the optimized regulators are still trivially performed. Evaluating the
flow equation (\ref{eq:flow2ptfn}) one obtains after analytically
continuing $p_0 = -\imag \omega$, and with the same notations 
 $\Gamma^{(0,2)}(\omega ) \equiv
\Gamma^{(0,2)}(p=(-\imag \omega,\vec0);\phi)$ as in
Sec.~\ref{sec:polemass} above, 
\begin{widetext}
\begin{equation}
\label{eq:flowequation2ptfnexplicit}
\begin{split}
\partial_t
\Gamma^{(0,2)}_{k,\pi\pi}(\omega )=\frac{k^5}{6\pi^2}\Biggl(&-\frac{(N+1)
  U_k''}{{E_k^\pi}^3}+
\frac{2 U_k''({E_k^\sigma}^2-{E_k^\pi}^2)\left((E_k^\sigma+E_k^\pi)^3
    ({E_k^\pi}^2+ E_k^\sigma
    E_k^\pi+{E_k^\sigma}^2)-({E_k^\sigma}^3+{E_k^\pi}^3)
    \omega^2\right)}{{E_k^\pi}^3 {E_k^\sigma}^3 
     \left((E_k^\pi+E_k^\sigma)^2-\omega^2\right)^2}\\
&-\frac{U_k''+2\phi^2 U_k^{(3)}}{{E_k^\sigma}^3}
 +\frac{8 N_f N_c g^2 (4 \epsilon_k^2+ \omega^2)}{\epsilon_k 
 \left(4 \epsilon_k^2- \omega^2\right)^2}\Biggr)\  .
\raisetag{\baselineskip}
\end{split}
\end{equation}
\end{widetext}
We set $\Gamma^{(0,2)}_{\Lambda,\pi\pi}(\omega)\equiv-\omega^2+2 U_{k=\Lambda}'(\phi^2)=  -\omega^2 +
\lambda (\phi^2 - v^2)$ at the
UV scale $k=\Lambda$, and obtain the pion pole mass
$m_{\pi,\text{pole}}$ for $k\to 0$ from the condition 
\begin{equation} 
\label{eq:defpolemass}
 \Gamma^{(0,2)}_{k=0,\pi\pi}(m_{\pi,\text{pole}}^2)\, =\, 0 \, ,
\end{equation}
evaluated at the minimum of the full effective potential.
For vanishing external momentum the two-point function can equally
be obtained from the second derivative of the effective potential.
Indeed, one verifies that the flow equation
(\ref{eq:flowequation2ptfnexplicit}) obeys the consistency condition 
\begin{equation}
\label{eq:2ptconsistency}
\partial_t \Gamma^{(0,2)}_{k,\pi\pi}(0)= \frac{\delta_{ij}}{N\!-1} \,
\frac{\partial^2}{\partial{\pi_i}\partial{\pi_j}}\partial_t U_k=2\frac{\partial}{\partial \phi^2}\partial_t U_k.
\end{equation}
This implies that if we calculate $\Gamma^{(0,2)}_{k=0,\pi\pi}$ by
integrating the flow equation (\ref{eq:flowequation2ptfnexplicit})
for $\omega =0$, the mass defined as
\begin{equation}
\label{eq:defscreenmass}
{m_\pi^\mathrm{sc}}^2 =  \Gamma^{(0,2)}_{k=0,\pi\pi}(0)
\end{equation}
correspondingly, simply represents the same screening mass as obtained  
from the curvature of the effective potential at its minimum, which is
usually considered in QM model calculations within the FRG framework.   

The flow equation (\ref{eq:flowequation2ptfnexplicit}) can be solved
via a Taylor expansion method around a scale dependent expansion point
for both the effective potential and the two-point function, or on a
grid in field space. In order to maintain the relation in 
\Eq{eq:2ptconsistency} also in the numerical calculations based on
Taylor expansions in $\phi^2$, one has to use one expansion order less for the
2-point function than for the effective potential. In this way we
can compute an estimate of the pion pole mass from a given UV
potential. % at $T=\mu=0$. 

\begin{table}[hb]
\begin{tabular}{|c||c|c|}
\hline method& quantity& value[MeV] \\ 
\hline 
Grid & $f_\pi$ & 76.0 \\
  & $m_{\pi,\text{scr}}$ & 178.8 \\
  & $m_{\sigma,\text{scr}}$ & 551.7 \\
  & 2$\mu_c$& \textbf{137.8}\\
  & $m_{\pi,\text{pole}}$ & \textbf{122.45} \\
  & $m_{\pi,\text{pole, ferm. only}}$ & 124.9 \\
    & $m_{\pi,\text{pole, bos. only}}$ & 171.6 \\
\hline
Taylor & $f_\pi$ & 75.0\\
	& $m_{\pi,\text{scr}}$ & 180.0 \\
  & $m_{\sigma,\text{scr}}$ & 550.8 \\
  & $m_{\pi,\text{pole}}$ & \textbf{122.6} \\
  & $m_{\pi,\text{pole, ferm. only}}$ & 125.0 \\
    & $m_{\pi,\text{pole, bos. only}}$ & 172.6 \\
    \hline
\end{tabular} 
\caption{Comparison of RG screening vs.~pole masses; `ferm.~only' (`bos.~only') refers to maintaining only the constant $\omega=0$ contributions in the bosonic (fermionic) contribution to the flow of the pion 2-point function, \Eq{eq:flowequation2ptfnexplicit}.}  
\label{tab:rgmasscomparison}
\end{table}

Table~\ref{tab:rgmasscomparison} shows a comparison of screening and
pole masses as obtained from the Taylor and grid methods. All
calculations here were performed at $T=\mu=0$. As explained in
Sec.~\ref{sec:polemass} and Appendix~\ref{sec:paramfixing}, we have
adjusted the start parameters for the flow in our two-dimensional 
grid code to fix the onset of diquark condensation to occur at $2
\mu_c\approx 138$ MeV which defines the physical pion mass in the
normal phase. The exact same parameters were used to obtain the UV
forms of effective potential and inverse propagators for the
one-dimensional Taylor expansion method at $\mu=0$. The results
from one- and two-dimensional grid computations at $\mu=0$ are
indistinguishable at this level of accuracy, as are the screening
masses from \Eq{eq:defscreenmass} and from the effective
potential. The slight deviations in $f_\pi$ and the masses in Table
\ref{tab:rgmasscomparison} between the grid and Taylor methods are an
indication of the small residual uncertainties. 

With the onset at half the physical pion mass fixed, we then observe
that the standard screening masses generally overestimate the pion
mass by about 30\%. In contrast, our pion pole mass estimates based on solving  
Eqs.~(\ref{eq:flowequation2ptfnexplicit}) and (\ref{eq:defpolemass})
lie within  11\%, but they are smaller than the physical one.

\begin{figure}[t]
\centering
\includegraphics[width=0.48\textwidth]{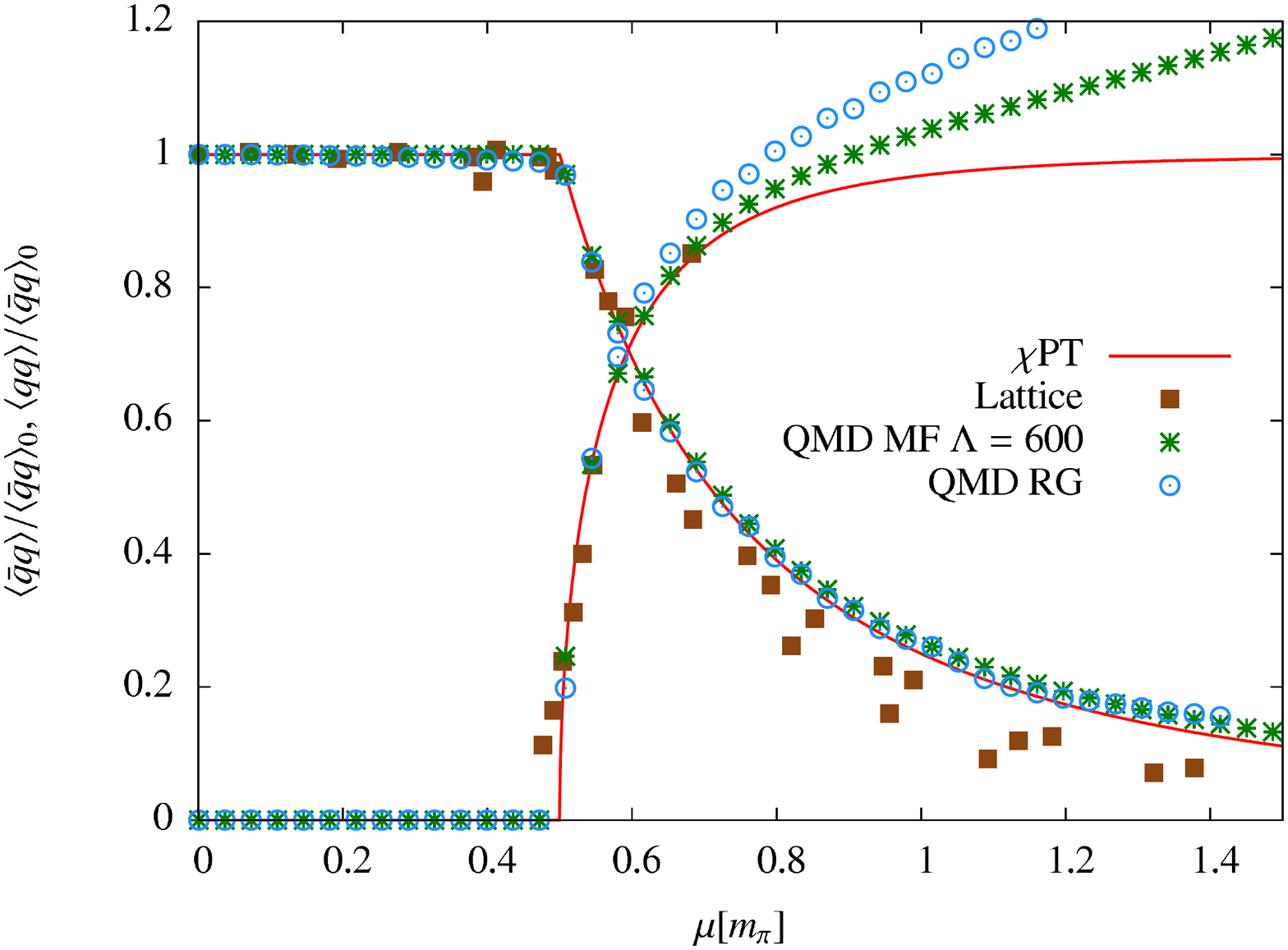}
\vspace{-.2cm}
\caption{Zero temperature condensates from full flow compared to 
  mean-field results (and the lattice data from \cite{Hands:2000ei}).
\vspace{-.2cm} }
\label{fig:t0axisrg}
\end{figure}

The extrapolation from zero pion momentum in the leading order derivative
expansion to the pion pole in the propagator from our consistent truncation
scheme appears to be too large, so that it overcompensates the
difference between onset and screening mass. In order to disentangle
the effect of bosonic and fermionic contributions to the flow equation 
(\ref{eq:flowequation2ptfnexplicit}) for the pion 2-point function, we
have also solved this equation  with $\omega = 0$ in the bosonic  
and in the fermionic parts, separately. The resulting pole masses 
are denoted by  $m_{\pi,\text{pole, ferm. only}}$ and $
m_{\pi,\text{pole, bos. only}}$ in Table~\ref{tab:rgmasscomparison},
respectively. Both contributions reduce the screening masses, but the
fermions clearly generate the dominant effect. This suggests that one
might have to go beyond the leading-order derivative expansion
%local potential approximation 
employed here and allow for an RG flow of the Yukawa couplings by including  
field renormalizations and anomalous dimensions 
%at least for the bosonic fields 
\cite{Braun:2009si}.

%\subsection{Phase diagram including diquark fluctuations and comparison to mean-field}

\subsection{Phase diagram of the QMD model for two-color
 QCD with mesonic and baryonic fluctuations}

In Fig.~\ref{fig:t0axisrg} we show once more the dependence of the chiral and
diquark condensates on the chemical potential at zero temperature as
in Fig.~\ref{fig:t0axismf}, but this time with including our results
from the full RG solution to \Eq{eq:fullflowfinal} on a
two-dimensional grid in field space with $\Phi =1$. 

\begin{figure}[t]
\centering
\includegraphics[width=0.48\textwidth]{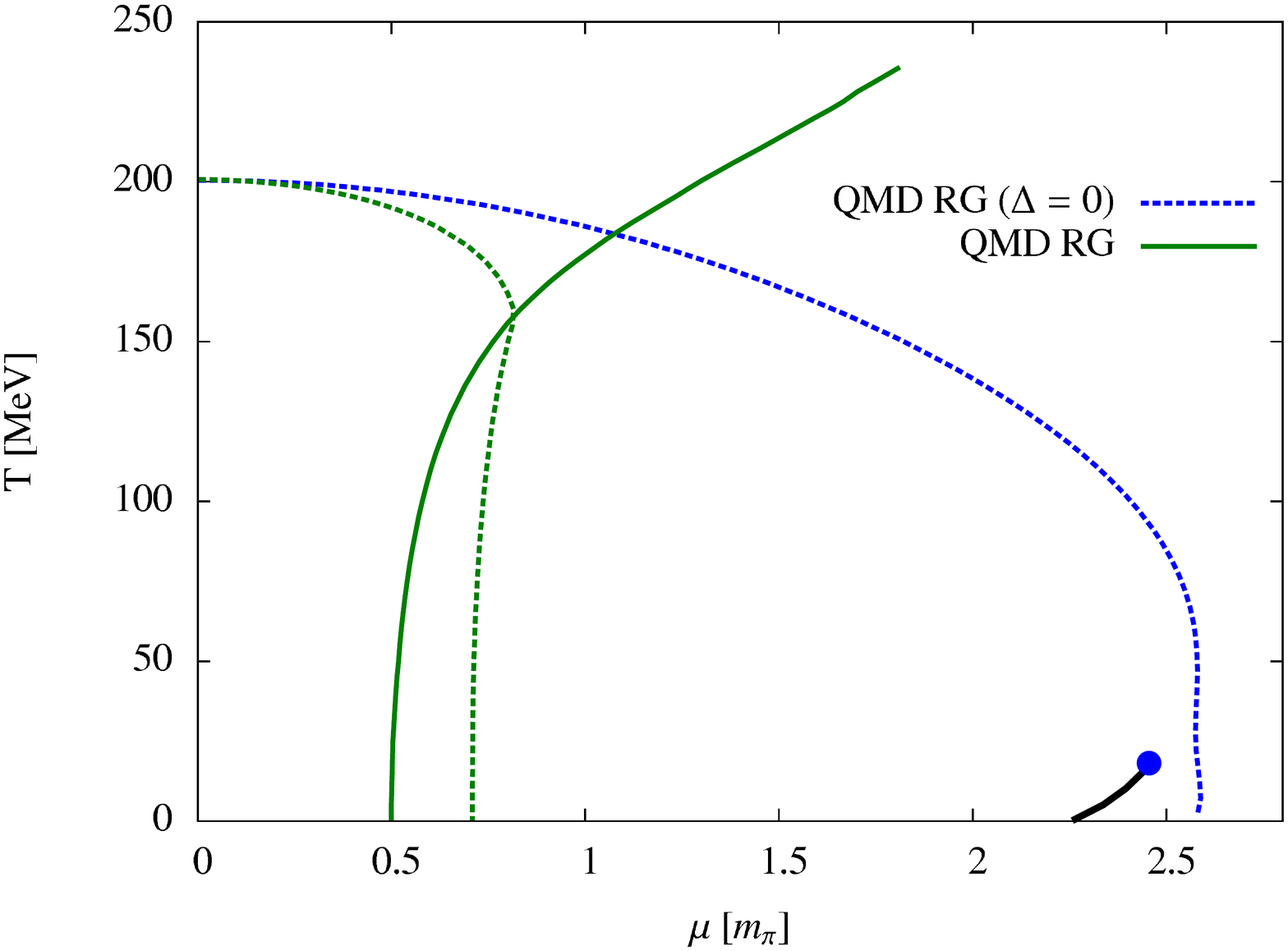}
\vspace{-.2cm}
\caption{Phase diagram from RG flow with collective baryonic
  fluctuations (and no chiral $1^\mathrm{st}$
  order transition/critical endpoint) compared to the purely mesonic
  model ($\Delta=0$). 
%\vspace{-.2cm} 
}
\label{fig:pdrg}
\end{figure}

The final effect of baryonic-diquark degrees of freedom is illustrated in
Fig.~\ref{fig:pdrg} where we compare the phase diagram from 
the one-dimensional RG flow solution to the $SO(6)$-symmetric equation
(\ref{eq:floweqdelta0}) from Sec.~\ref{sec:pwwodq} and that from the
full two-dimensional one for an effective potential with the reduced
$SO(4)\times SO(2)$ symmetry.

This clearly illustrates the effect of the competing dynamics of the
collective mesonic and baryonic fluctuations. As before, the dashed
lines in Fig.~\ref{fig:pdrg} indicate the chiral crossover by tracing the
half-value of the chiral condensate. Both, the one and the
two-dimensional results agree for quark-chemical potentials near
zero. The crossover in this region leads to mesonic freeze-out as
usual, and the results are unambigously determined by the $O(6)$
symmetry breaking pattern, see Sec.~\ref{sec:O6scale}. Allowing
additional interactions with lower symmetry has no effect on the flow
here.
 
Once the quark-chemical potential approaches half the baryon mass,
corresponding to $m_B/N_c $, however, the rapidly increasing
baryon density equally rapidly suppresses the chiral condensate. With
the proper inclusion of the collective baryonic excitations, there is
no trace left of the chiral first-order transition and the critical
endpoint of the purely mesonic model. The baryon density is an order
parameter for $N_f=N_c=2$, and the transition line would be expected
to give rise to the two-color analogue of the baryonic freeze-out.

\begin{figure}[t]
\centering
\includegraphics[width=0.48\textwidth]{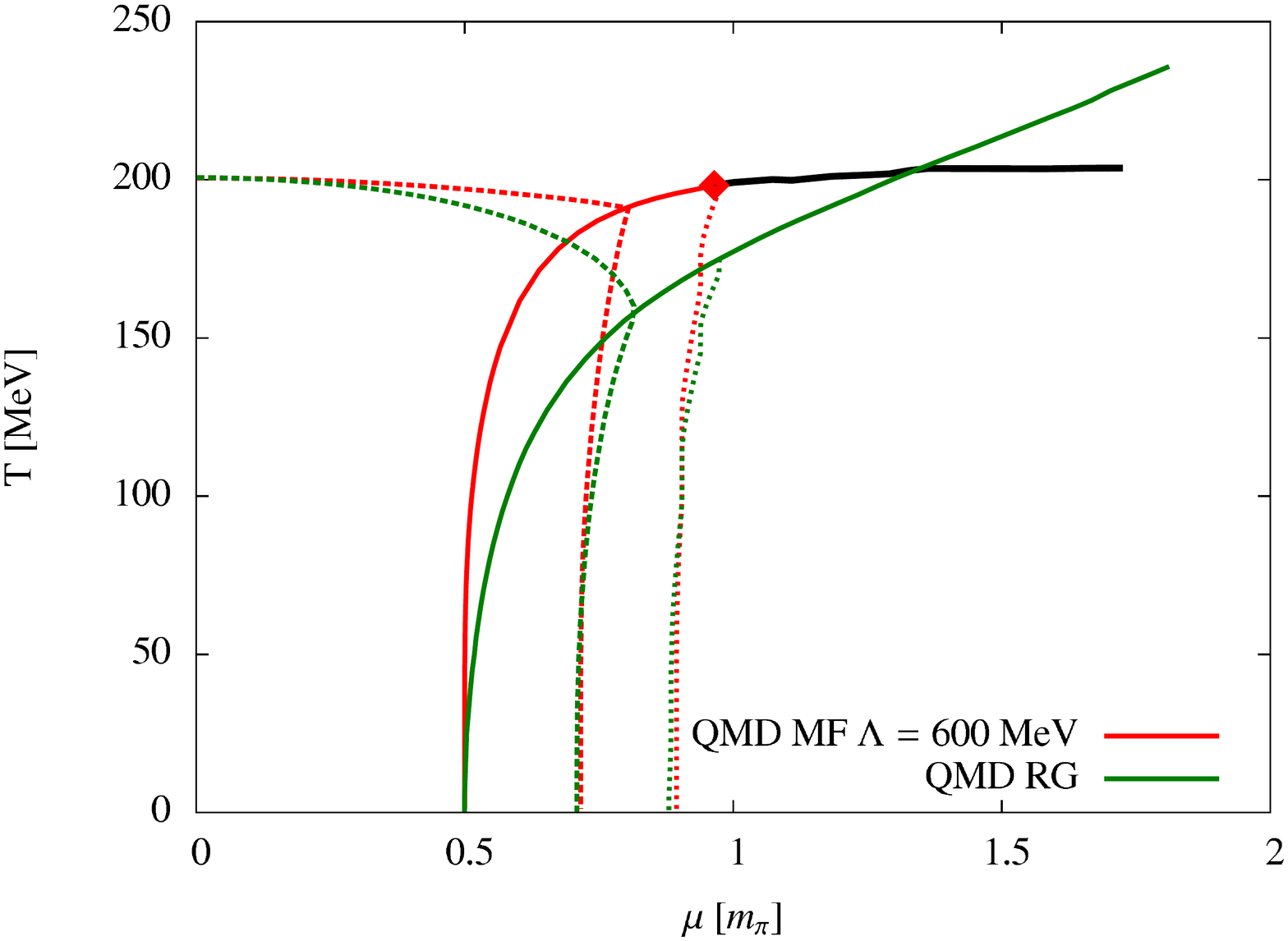}
\vspace{-.2cm}
\caption{Comparison of QMD phase diagrams from MF and RG calculations,
  including lines with $g\sigma = \mu$ in the superfluid phase to
  indicate the BEC-BCS crossover. 
\vspace{-.5cm} }
\label{fig:pdrg2}
\end{figure}

The onset of diquark condensation and superfluidity of our bosonic
baryons, with $SO(3)\times SO(2) \to SO(3)$ symmetry breaking at finite
quark mass and chemical potential, also marks the line at which the   
the residual $SO(3)$ symmetry starts changing in nature
from an approximate $SO(5)$ symmetry as in the normal phase to
becoming the approximate  $SO(4) \simeq SU(2)_L\times SU(2)_R$
quasi-restored chiral symmetry. Because they are both explicitly
broken and only approximate symmetries, this vacuum realignment 
naturally is a crossover. The quark mass with large chiral condensate in the
normal phase starts out as a predominantly spontaneously generated
Dirac mass, and the bosonic baryons undergo Bose-Einstein condensation
as a dilute gas of strongly bound diquarks with the onset of diquark
superfluidity.   
As their density increases, the underlying quark mass
rotates into a spontaneous Majorana mass leading to a BCS-like
pairing. This is the relativistic analogue in two-color QCD 
of the BEC-BCS crossover observed in ultracold fermionic quantum
gases. It is indicated in  Fig.~\ref{fig:pdrg2} as additional dashed
lines in the superfluid phase tracing the lines where the quarks'
Dirac-mass $m_q = g\sigma$  equals their chemical potential, {\it
  i.e.} $\mu = m_q$, see \cite{He:2010nb} for a comprehensive
discussion of this crossover within the NJL model.
   
 %Comparison
In this Figure \ref{fig:pdrg2} we compare the phase diagram of the QMD
model for two-color QCD as obtained from the full RG solution with the 
mean-field result of Sec.~\ref{sec:DiqCond}.
The line of the diquark-condensation phase transition, which  
one expects to be of  $O(2)$-universality,
in the QMD model RG solution with fluctuations differs more and more
from that obtained in mean-field QMD and NJL model calculations as
temperature increases. 
The first-order transition line is washed out by the fluctuations and the  
associated tricritical point as also predicted from next-to-leading
order $\chi$PT \cite{Splittorff:2001fy} turns out to be a mean-field
artifact. As already visible from the $T=0$ results for the condensates,
{\it c.f.} Fig.~\ref{fig:t0axisrg}, the phase diagrams approach one
another at small temperatures.

\section{Summary and Outlook\label{sec:outlook}}

\begin{figure}[ht]
\centering
\includegraphics[width=0.48\textwidth]{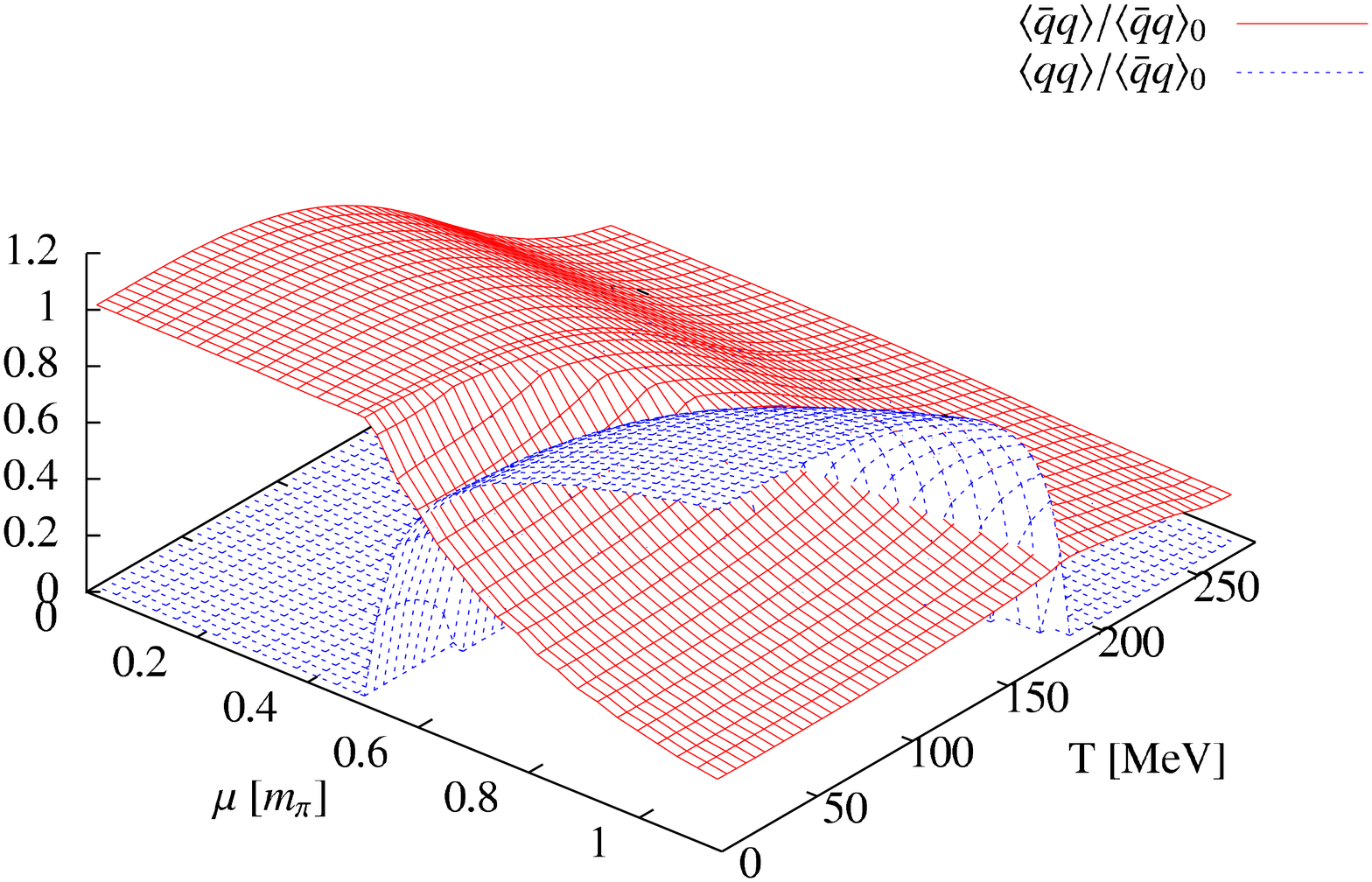}
\vspace{-.2cm}
\caption{Summary: Chiral condensate and diquark condensate as function of temperature and chemical potential from RG calculation.
\vspace{-.2cm} }
\label{fig:pd3d}
\end{figure}

In this paper we have developed a Polyakov-loop extended
quark-meson-diquark model for two-color QCD and derived the functional
renormalization group equation for the grand potential in the
leading-order derivative expansion. We discussed the mean-field
thermodynamics of the model and solved the RG flow equation for
trivial Polyakov-loop, {\it i.e.} for the corresponding
quark-meson-diquark model. In order to correctly describe the
competing dynamics of collective mesonic and baryonic diquark
fluctuations, it is thereby necessary to introduce two invariants of
the fields in order to account for the rich symmetry and
symmetry-breaking structure of two-color QCD as reviewed in our
introduction. The functional RG for the effective potential
then describes the interplay between the collective mesonic and
baryonic (diquark) fluctuations as summarized once more    
with showing the resulting chiral and diquark condensates over
temperature and quark chemical potential in a three-dimensional plot in
Fig.~\ref{fig:pd3d}. 
Our numerical solution method on a higher-dimensional grid
in field space %, here including chiral as well as diquark condensate,
represents important technical progress with many further
applications.         

One particular advantage of using two instead of the usual three
colors is that our non-perturbative functional methods and model
results can be tested against exact results and lattice simulations in
two-color QCD. First important results from such tests are: The
expected $O(6)$ scaling at zero density; the relevance of pole masses
in the RG framework to correctly describe the onset of diquark
condensation at the zero temperature quantum phase transition of
two-color QCD, and the failure of the usual screening masses to be
capable of that; and finally but most importantly, the non-existence
of a chiral first-order transition and critical endpoint at finite
baryon density. 

The latter is not surprising for two-color QCD alone, with the
BEC-BCS crossover in the superfluid phase of the bosonic baryons. We
argue, however, that our comparison between the full results with
inclusion of collective baryonic excitations and the corresponding
purely mesonic model reveals a general effect, relevant to the real
world: the chiral condensate drops discontinuously at the low
temperature liquid-gas transition to nuclear matter, and it will
continue to decrease with increasing baryon density so that one might
question whether there will be enough chiral-symmetry breaking left
for a another first-order transition at the expected higher
densities. Similarly, one might speculate that the second-order
phase-transition line of two-color QCD to diquark superfluidity at
finite temperature would lead to the analogue of the observed baryonic
freeze-out line in the region of rapidly increasing baryon density in
real QCD. % with three colors. 

Another advantage is that the proper inclusion of baryonic degrees of
freedom is much more straightforward and much simpler here as in real
QCD. While this is to a large extent due to the fact that those
baryons are represented by bosonic diquarks, our study of two-color QCD
can serve as an important first step towards including
diquark-correlations and explicit baryonic degrees of freedom in a
covariant quark-diquark description by a corresponding
quark-meson-baryon model for QCD.    

More tests of refined truncations will be performed in the future,
including a dynamical coupling of the quark-meson-diquark model
studied here to the full gauge-field dynamics of two-color 
QCD along the lines of what has been done already without explicit
baryonic contributions for QCD \cite{arXiv:0908.0008,Pawlowski:2010ht}.

\acknowledgments The authors thank Tom\'{a}\v{s} Brauner, Michael Buballa,
Holger Gies, Kazuhiko Kamikado, Mario Mitter, Jan Pawlowski, David Scheffler
and Jochen Wambach for useful discussions. This work was supported 
by the Helmholtz International Center for FAIR within the LOEWE program 
of the State of Hesse, the Helmholtz Association Grant VH-NG-332, and 
the European Commission, FP7-PEOPLE-2009-RG No. 249203.

\newpage
%%%%%%%%%%%%%%%%%%%%%%%%%%%%%%%%%%%%%%%%%%%%%%%%%%%%%%%%%%%%%%%%%%%%%%%%%%%%%
\appendix
%%%%%%%%%%%%%%%%%%%%%%%%%%%%%%%%%%%%%%%%%%%%%%%%%%%%%%%%%%%%%%%%%%%%%%%%%%%%%
\section{Parameter fixing and numerical procedure\label{sec:paramfixing}}

In this appendix we briefly outline our parameter fitting and the
numerical methods for the solution of the flow equation.

Since two-color QCD is an unrealistic theory there is no canonical
choice to fix the parameters to measurable quantities. A common
approach, often found in the literature, is to use the experimentally
known $N_c=3$ values and their $N_c$ scaling to obtain a consistent
$N_c=2$ parameter set. Therefore the assumption $f_\pi\sim \sqrt{N_c}$
yields $f_\pi=76$ MeV if the usual three-color value
$f^{(3)}_\pi=93$ MeV is chosen.
%  we assume $f_\pi\sim \sqrt{N_c}$
% which leads to $f_\pi=76$ MeV instead of the ususal 3-color value of
% 93 MeV. 
Furthermore, we assume that the vacuum pion and sigma masses do not scale
with $N_c$ and fix the pion mass to $m_\pi= m_{\pi,0}= 138$ MeV.

As pointed out above the mass definition at mean-field level which is
consistent with the Silver Blaze property is the pole mass defined via
\Eq{eq:polnormal}. Usually, in QM studies we fix
$\sigma(T=\mu=0)=f_\pi$ together with the pion and the sigma masses in
the vacuum. The pion and sigma pole mass equations \Eq{eq:polnormal}
fix the constants $\lambda$ and $v^2$ in the
potential $V=\frac{\lambda}{4}(\phi^2-v^2)^2-c\sigma$. The explicit
symmetry breaking constant $c$ is then determined by the gap
equation. In this way the parameters $\lambda$, $v^2$ and $c$ are
found for a fixed momentum cutoff $\Lambda$ in the vacuum term. As
argued above the cutoff $\Lambda$ should be chosen larger than the
largest value of the chemical potential we are interested in. Here we
choose $\Lambda=600$ MeV and for comparison $\Lambda=0$ MeV and adjust
$m_\sigma$ such that the crossover temperature at $\mu=0$ coincides
with the RG calculation.  In Table \ref{tab:mfparams} we summarize our
used parameter values.
\begin{table}[ht]
\begin{tabular}{|c|c||c|c|c|c|}
\hline $\Lambda$[MeV] & $m_\sigma$[MeV]&$g$ & $\lambda$ &$v^2$ [$10^3$ MeV$^2$]& $c$ [$10^6$ MeV$^3$] \\ 
\hline 600 & 680 & 4.8 & 25.05 &-42.710&2.885 \\
\hline 0 & 1055 & 4.8 & 94.70 &5.575&1.447 \\
\hline 
\end{tabular} 
\caption{Parameter values in mean-field approximation}
\label{tab:mfparams}
\end{table}

In the RG setting we adjust the parameter $\lambda$ in the UV potential 
and the explicit symmetry breaking parameter $c$ while keeping $v^2=0$ 
in the UV potential and a Yukawa coupling $g=4.8$ to match $f_\pi$ and 
$m_\pi$ in the IR. As there are remaining uncertainties in the 
determination of the pion pole mass via the flow of 2-point function as explained 
in Sec.~\ref{sec:flow2ptfn}, we rather determine $m_\pi$ via the onset 
of diquark condensation at $T=0$.

Finally, we point out some details on the numerical procedure to solve
the flow equations. The structure of the $d=0$ flow
(\ref{eq:floweqdelta0}) and the full flow for $\mu=0$,
\Eq{eq:flowmu0}, is identical to the flow of the usual three-color QM
model \cite{Bohr:2000gp, Schaefer:2004en}.  Several solution methods
such as the finite difference approach, the Taylor expansion of the
effective potential around a scale dependent minimum or grid
techniques where higher order derivative terms are eliminated
algebraically are known which all produce consistent results. For
the full flow equation (\ref{eq:fullflowfinal}) the situation is more
involved since the effective potential is parametrized by two
invariants $U_k=U_k(\rho^2,d^2)$.

Here we apply a modified grid algorithm where the higher derivatives on
the RHS of the flow equation are obtained by a two-dimensional-spline
fit of the effective potential at the respective grid
point. The numerical results obtained from this procedure agree very well
with those from an algorithm where the derivatives 
were approximated by finite differences at a fixed
discretization order.

\section{Proper-time flow equations\label{sec:ptRG}}

In this appendix we sketch an alternative derivation of the QMD flow
equation \Eq{eq:fullflowfinal} with a proper-time regularization, see
also \cite{Litim:2010tt, Schaefer:2006sr, Schaefer:1999em} and
references therein for a short introduction and comparison of the
proper-time with the Wetterich flow.

Originally, the proper-time renormalization group equation (PTRG) was
found by an RG improvement of a proper-time regularized one-loop
effective action \cite{Liao:1994fp, Schaefer:1999em}. Later, it turned
out \cite{Litim:2002hj} that the PTRG flow can be related to the
Wetterich flow with the background field formalism. The standard PTRG
flow can be derived from the Wetterich flow when terms proportional to
$\partial_t \Gamma^{(2)}_k$ are neglected. 
The PTRG flow for the QMD model splits into a bosonic and fermionic
flow $\partial_t \Gamma_k=\partial_t \Gamma_{k,B}+\partial_t
\Gamma_{k,F}$ where $t=\log(k/\Lambda)$ denotes the logarithmic RG
scale. The one-loop expression can be rewritten via Schwinger's
proper-time representation as
\begin{equation}
\begin{split}
&\partial_t \Gamma_k= -\frac{1}{2} \Tr \int\limits_0^\infty
\frac{\d\tau}{\tau}\partial_t f_a(\tau k^2) \\
& \left[\exp\left(-\tau \Gamma_{k,B}^{(2)}\right) - \exp\left(-\tau
    \Gamma_{k,F}^{(2)}(\mu)\Gamma_{k,F}^{(2)\dagger}(-\mu)\right)\right]\ ,
\end{split} 
\end{equation}
where the trace runs over momenta and internal indices.\footnote{For
  the fermionic part we make use of $\gamma_5$ hermiticity and parity
  invariance  of the Dirac operator by writing (using the notation of
  \Eq{eq:S0-1}) $A^5 \Gamma_{F,k}^{(2)}(\sigma,\vec \pi,\Delta;\mu)
  A^5=\Gamma_{F,k}^{(2)\dagger}(\sigma,-\vec
  \pi,\Delta;-\mu)=\Gamma_{F,k}^{(2)\dagger}(\sigma,\vec
  \pi,\Delta;-\mu)$ with
  $A^5=\left(\begin{smallmatrix}\gamma^5&0\\0&-\gamma^5\end{smallmatrix}\right)$.} 
As before, the second functional 
derivative of effective action with respect to bosonic(fermionic) fields is denoted by $\Gamma_{k,B}^{(2)}$ $(\Gamma_{k,F}^{(2)})$. The
proper-time regulator function $f_a(\tau k^2)$ has to fulfill some
constrains and the optimal choice, based on incomplete Gamma
functions, is $f_a(\tau k^2)={\Gamma(a+1,\tau k^2)}/{\Gamma(a+1)}$
with $a=3/2$ which can also be mapped to the optimized regulator in
the Wetterich flow, for details see
\cite{Litim:2001hk}. 

In the bosonic case $\Gamma^{(2)}_{k,B}$ is given by \Eq{eq:gamma2b}
and can be diagonalized. Three of the six eigenvalues which are
related to the three massless pions are degenerate and read
explicitly $\vec p^2 +\lambda_{n,k}^{(i)} =\vec p^2 +\omega_n^2+2\Ur$,
$i=1,2,3$ where $\omega_n = 2\pi n T$ are the bosonic Matsubara
frequencies. The remaining three eigenvalues $\vec p^2
+\lambda_{n,k}^{(i)}$, $i=4,5,6$ are more complicated and are related
to the radial $\sigma$-meson and the two diquarks.  The three-momentum 
integration separates and can be done analytically. After
the proper-time integration the bosonic flow is composed of a sum over
all eigenvalues
% \begin{equation}
% \begin{split}
% \partial_t \Gamma_{k,B} &=\frac{1}{2}\sum_{i=1}^6\int_0^\infty\frac{\d\tau}{\tau} T\sum_n\int\frac{\d^3 p}{(2\pi)^3}  \frac{8}{3\sqrt{\pi}}(\tau k^2)^{5/2}\exp(-\tau (k^2+\lambda_i))\\
% &=\frac{T}{3}\frac{k^5}{2\pi^2}\sum_{i=1}^6\sum_n\int_0^\infty \d\tau \exp(-\tau (\lambda_i(\vec p^2\to k^2)))\\
% &=\frac{T}{3}\frac{k^5}{2\pi^2}\sum_{i=1}^6\sum_n \frac{1}{k^2+\lambda_i(\vec p^2\to k^2)}
% \end{split}
% \end{equation}

\vspace{-.5cm}

\begin{equation}
\partial_t \Gamma_{k,B}
=\frac{T}{3}\frac{k^5}{2\pi^2}\sum_{i=1}^6\sum_{n \in Z}
\frac{1}{k^2+\lambda_{n,k}^{(i)}}\ .
\vspace{-.5cm}
\end{equation}
Rewriting

\vspace{-.5cm}

\begin{equation}
\sum\limits_{i=4}^6 \frac{1}{k^2 + \lambda_{n,k}^{(i)}}  = 
\frac{\alpha_2 (\omega_n^2)^2+\alpha_1   \omega_n^2+\alpha_0}
{(\omega_n^2)^3+ \beta_2 (\omega_n^2)^2
  +\beta_1 \omega_n^2 +\beta_0}
\end{equation}
with the $k$- and $\mu$-dependent coefficient functions $\alpha_{l}$ and
$\beta_{l}$ listed explicitly in Appendix~\ref{sec:bosonicflowcoefficients} we arrive at the bosonic flow equation \Eq{eq:bosonicflow} again.

In the fermionic sector including the coupling to the gauge field via
the Polyakov loop variable $\Phi=\cos (\beta a_0)$ we find two $4
N_f(=8)$-fold degenerate eigenvalues $\lambda^\pm_{n,k}$ of the
matrix $\Gamma^{(2)}_{k,F}(\mu) \Gamma^{(2)\dagger}_{k,F} (-\mu)$
\begin{equation}
\label{eq:eigenvaluesfermioncflow}
\lambda^\pm_{n,k} = (\nu_n+a_0)^2-\mu^2+g^2\phi^2\pm 2
\imag\mu\sqrt{(\nu_n+a_0)^2+g^2|\Delta|^2}\ ,
\end{equation}
with the fermionic Matsubara frequencies $\nu_n = (2n+1) \pi T$. This
thus yields the fermionic flow of the effective potential

%\parbox{\linewidth}{
%\vspace*{-.5cm}
\begin{equation}
\begin{split}
\label{eq:fermionicflowfinal}
&\partial_t \Gamma_{k,F}= -\frac{4 T
  k^5}{3\pi^2}\sum_{j=\pm}\sum_{n\in Z} \frac{1}{k^2+\lambda^j_{n,k}} \\
&=-\frac{8k^5 T}{3\pi^2}\!\sum_{n \in Z}\!
  \frac{k^2+g^2\phi^2  -\mu^2+(\nu_n+a_0)^2}
  { \left((\nu_n+a_0)^2+{E_k^+}^2\right)\!\left(
    (\nu_n+a_0)^2+{E_k^-}^2\right) },
\end{split}
\end{equation}
%\vspace{.5cm}
%}
which reproduces \Eq{eq:fermionicflow}.  Evaluating the Matsubara sums and
combining both contributions then leads to the flow equation
(\ref{eq:fullflowfinal}).

%%%%%%%%%%%%%%%%%%%%%%%%%%%%%%%%%%%%%%%%%%%%%%%%%%%%%%%%%%%%%%%%%%%%%%%%%%%%%
\section{RPA Meson/Diquark polarization functions\label{sec:rpamass}}

For convenience we indicate the explicit expressions for the
meson/diquark polarization functions for vanishing spatial external
momentum.  
These can be calculated most conveniently using massive energy
projectors \cite{He:2005nk}. 
The Polyakov-loop enhanced quark/antiquark occupation numbers $N_q$ are defined
in \Eq{eq:PLenhancedoccnumbers} and reduce to the Fermi Dirac
distribution for $\Phi=1$. 
To comply with conventions in the literature the polarization functions
are given in a complex basis $\phi=(\sigma,\vec \pi,\Delta,\Delta^*)$
and correspondingly  
with \Eq{eq:rpadefinition} replaced by
\begin{equation}
\Pi_{ij}(p)=\Tr_q\left[ \frac{\partial \Gamma^{(2)}_F}{\partial \phi^*_i}\Big|_{\phi_{\mathrm{MF}}}\hspace{-1mm} G_{\mathrm{MF}}(p+q) \frac{\partial \Gamma^{(2)}_F}{\partial \phi_j}\Big|_{\phi_{\mathrm{MF}}}\hspace{-1mm} G_{\mathrm{MF}}(q)\right].
\end{equation}
%Note that $\Pi_{\sigma\sigma}$ given below contains an additional
%contribution proportional to $({E_p^\pm}^2-{\epsilon_p^\pm}^2)$ which
%only contributes 
%in the diquark condensation phase and was left out in
%\cite{He:2005nk,Xiong:2009zz}.  
There is a subtlety concerning the $\omega\to 0$ limit at
finite temperature $T>0$ as mentioned in Sec.~\ref{sec:polemass}. 
The standard procedure within the imaginary-time formalism
assumes that the external Euclidean $p_0 = -\imag \omega$ is a discrete
Matsubara frequency $2\pi n T$ ($n\in\mathbb{Z}$). One then assumes
additional analyticity properties to define a unique analytic
continuation. The polarization functions are then singular in the
origin of momentum space and one needs to maintain a finite spatial 
external momentum $\vec p$ to define dynamic sceening masses via the
limit $|\vec p|\to 0$ at $\omega = 0$. This leads to a discontinuity
at $\omega = 0$ and gives rise to additional
contributions $\delta_{\omega, 0} \Pi_{ij}^0$ for the zero mode
$\omega=0$ (see below). Note that 
polarization functions for Goldstone and would-be-Goldstone modes 
are protected from these contributions $\delta_{\omega,0}
\Pi^0 $, \textit{ i.e.}, only $\Pi_{\sigma\sigma}^0$ is nonzero in the
normal phase. Using  
$m_q=g\sigma$ and $\epsilon_q= \sqrt{
  \vec{q}^2 + m_q^2}$, $\epsilon_q^\pm = \epsilon_q\pm \mu$,
$E_q^\pm = \sqrt{
  {\epsilon_q^\pm}^2  + g^2d^2} $ as in
\Eqs{eq:MFeigenvalues} the polarization functions are given
by\footnote{Note that apart from the missing zero-mode contributions
  $\delta_{\omega, 0} \Pi_{ij}^0$, the prefactor of   
the terms proportional to $m_q^2$ in $\Pi_{\sigma\sigma} $ in the last line
of \Eq{eq:C3} differs by a factor of 2 from the corresponding terms
given in \cite{He:2005nk,Xiong:2009zz}.} 

\begin{widetext}
\begin{align}
\label{eq:C2}
\Pi_{\pi_i\pi_j}(\omega,T)=&\,-4 N_c g^2 \delta_{ij} \int\frac{\d^3
  q}{(2\pi)^3}\Biggl[\frac{E_q^+ E_q^- - \epsilon_q^+\epsilon_q^- -g^2
  d^2}{\omega^2-(E_q^--E_q^+)^2}\left(\frac{1}{E_q^+}-\frac{1}{E_q^-}\right)\left(N_q(E_q^-)-N_q(E_q^+)\right)\notag\\ 
&-\frac{E_q^+ E_q^- + \epsilon_q^+\epsilon_q^-+g^2
  d^2}{\omega^2-(E_q^-+E_q^+)^2}\left(\frac{1}{E_q^+}+\frac{1}{E_q^-}\right)\left(1-N_q(E_q^-)-N_q(E_q^+)\right)\Biggr],\\
\Pi_{\sigma\sigma}(\omega,T)=&\, -4 N_c g^2 \delta_{ij} \int\frac{\d^3
  q}{(2\pi)^3}\frac{\vec q^2}{\epsilon_q^2}\Biggl[\frac{E_q^+ E_q^- - \epsilon_q^+\epsilon_q^- -g^2
  d^2}{\omega^2-(E_q^--E_q^+)^2}\left(\frac{1}{E_q^+}-\frac{1}{E_q^-}\right)\left(N_q(E_q^-)-N_q(E_q^+)\right)\notag\\ 
&-\frac{E_q^+ E_q^- + \epsilon_q^+\epsilon_q^-+g^2
  d^2}{\omega^2-(E_q^-+E_q^+)^2}\left(\frac{1}{E_q^+}+\frac{1}{E_q^-}\right)\left(1-N_q(E_q^-)-N_q(E_q^+)\right)\Biggr]\notag\\
  &+4 N_c g^2 \int\frac{\d^3
    q}{(2\pi)^3}\frac{m_q^2}{\epsilon_q^2}\sum_\pm
%\Biggl[\frac{g^2
%d^2+({E_q^\pm}^2-{\epsilon_q^\pm}^2)}{\omega^2-4{E_q^\pm}^2}
\Biggl[\frac{2g^2 d^2}{\omega^2-4{E_q^\pm}^2} 
\frac{1}{E_q^\pm}\left(1-2N_q(E_q^\pm)\right)\Biggr]+
  \delta_{\omega,0} \Pi^0_{\sigma\sigma}(T),\label{eq:C3}\\
\Pi_{\Delta\Delta}(\omega,T)=&\,\Pi_{\Delta^*\Delta^*}(-\omega,T)=4
N_c g^2 \int\frac{\d^3 q}{(2\pi)^3}\sum_\pm\Biggl[\frac{ {E_q^\pm}^2
  +{\epsilon_q^\pm}^2\mp\omega{\epsilon_q^\pm}}{\omega^2-4{E_q^\pm}^2}\frac{1}{E_q^\pm}\left(1-2
  N_q(E_q^\pm)\right)\Biggr]+
  \delta_{\omega,0} \Pi_{\Delta\Delta}^0(T),\\ 
\Pi_{\Delta\Delta^*}(\omega,T)=&\,\Pi_{\Delta^*\Delta}(\omega,T)=-4
N_c g^2 \int\frac{\d^3 q}{(2\pi)^3}\sum_\pm\Biggl[\frac{g^2
  d^2}{\omega^2-4 {E_q^\pm}^2}\frac{1}{E_q^\pm}(1-2
N_q(E_q^\pm))\Biggr]+   \delta_{\omega,0} \Pi_{\Delta\Delta^*}^0(T),\\ 
\Pi_{\sigma
  \Delta}(\omega,T)=&\,\Pi_{\Delta\sigma}(\omega,T)=\Pi_{\sigma
  \Delta^*}(-\omega,T)=\Pi_{\Delta^*\sigma }(-\omega,T)\notag\\ 
=&\,2\sqrt{2}N_c g^2 \int\frac{\d^3 q}{(2\pi)^3}\frac{ g d\,
  m_q}{\epsilon_q}\sum_\pm\Biggl[\frac{2
  \epsilon_q^\pm\pm\omega}{\omega^2-4
  {E_q^\pm}^2}\frac{1}{E_q^\pm}\left(1-2N_q(E_q^\pm)\right)\Biggr]+
   \delta_{\omega,0} \Pi_{\sigma\Delta}^0(T)  ,
\end{align}
with additional contributions for $\omega=0$ of the form,
\begin{align}
\Pi_{\sigma\sigma}^0(T)=&\,-g^2 N_c  \int\frac{\d^3 q}{(2\pi)^3}\sum_\pm
\frac{
  m_q^2}{\epsilon_q^2}\left(\frac{{E^\pm_q}^2+{\epsilon^\pm_q}^2-g^2
    d^2}{{E^\pm_q}^2}\right)\left(-2 N_q'(E_q^\pm)\right),\\ 
\Pi_{\Delta\Delta}^0(T)=&\,\Pi_{\Delta^*\Delta^*}^0(T)=-2 N_c g^2
\int\frac{\d^3 q}{(2\pi)^3}\sum_\pm
\left(\frac{{E^\pm_q}^2-{\epsilon^\pm_q}^2}{{E_q^\pm}^2}\right)\left(-2
  N_q'(E_q^\pm)\right),\\ 
\Pi_{\Delta\sigma}^0(T)=&\,\Pi_{\Delta^*\sigma}^0(T)=\Pi_{\sigma\Delta}^0(T)=\Pi_{\sigma\Delta^*}^0(T)=-2\sqrt{2}
N_c g^2  \int\frac{\d^3 q}{(2\pi)^3}\sum_\pm \frac{g
  d\, m_q}{\epsilon_q}\frac{\epsilon^\pm_q}{{E^\pm_q}^2}\left(-2
  N_q'(E_q^\pm)\right),\\ 
\Pi_{\Delta\Delta^*}^0(T)=&\,\Pi_{\Delta^*\Delta}^0(T)=-2 N_c g^2
\int\frac{\d^3 q}{(2\pi)^3}\sum_\pm \frac{g^2
  d^2}{{E^\pm_q}^2}\left(-2 N_q'(E_q^\pm)\right). 
\label{eq:C10}
\end{align}
\end{widetext}
The additional contributions $\Pi_0(T)$ vanish for $T\to 0$ but are
required to ensure consistency with the screening mass definition from
the effective potential at finite $T$, \textit{c.f.}
Eqs.~(\ref{eqs:Pizero}) and (\ref{eq:scrnormal}).

\section{Coefficients in the bosonic flow equation\label{sec:bosonicflowcoefficients}}
In this appendix we list the expressions for the coefficient functions $\alpha_i$ and $\beta_i$ appearing in \Eq{eq:fullflowfinal}.
\begin{widetext}
\begin{align}
\alpha_0 =&3 k^4 + 4 k^2 (-4 \mu^2 + 2 U_{k,d} + 2 d^2 U_{k,dd} + U_{k,\rho} + 2 \rho^2 U_{k,\rho\rho})\notag\\ 
&+ 4 \left(4 \mu^4 + U_{k,d}^2 + 2 U_{k,d} (d^2 U_{k,dd} + U_{k,\rho} + 2 \rho^2 U_{k,\rho\rho})- 4 \mu^2 (U_{k,d} + d^2 U_{k,dd} + U_{k,\rho} + 2 \rho^2 U_{k,\rho\rho})\right.\notag\\
&\left.\quad+ 2 d^2 (U_{k,dd} U_{k,\rho} - 2 \rho^2 U_{k,\rho d}^2 + 2 \rho^2 U_{k,dd} U_{k,\rho\rho})\right)\\
\alpha_1 =&6 k^2 + 8 U_{k,d} + 8 d^2 U_{k,dd} + 4 U_{k,\rho} + 8 \rho^2 U_{k,\rho\rho}\\
\alpha_2 =&3\\
\beta_0 =&(k^2 - 4 \mu^2 + 2 U_{k,d}) \left(k^4 + 2 k^2 (-2 \mu^2 + U_{k,d} + 2 d^2 U_{k,dd} + U_{k,\rho} + 2 \rho^2 U_{k,\rho\rho})\right.\notag\\ 
&\left. + 4 (-2 \mu^2 U_{k,\rho} + U_{k,d} U_{k,\rho} + 2 d^2 U_{k,dd} U_{k,\rho} - 4 d^2 \rho^2 U_{k,\rho d}^2 + 2 \rho^2 (-2 \mu^2 + U_{k,d} + 2 d^2 U_{k,dd}) U_{k,\rho\rho})\right)\\
\beta_1 =&3 k^4 + 4 k^2 (2 U_{k,d} + 2 d^2 U_{k,dd} + U_{k,\rho} + 2 \rho^2 U_{k,\rho\rho})\notag\\
 &+ 4 \left(4 \mu^4 + U_{k,d}^2 - 4 \mu^2 (U_{k,d} + d^2 U_{k,dd} - U_{k,\rho} - 2 \rho^2 U_{k,\rho\rho}) +  2 U_{k,d} (d^2 U_{k,dd} + U_{k,\rho} + 2 \rho^2 U_{k,\rho\rho})\right.\notag\\
 &\quad\left.+ 2 d^2 (U_{k,dd} U_{k,\rho} - 2 \rho^2 U_{k,\rho d}^2 + 2 \rho^2 U_{k,dd} U_{k,\rho\rho})\right)\\
\beta_2 =&3 k^2 + 8 \mu^2 + 4 U_{k,d} + 4 d^2 U_{k,dd} + 2 U_{k,\rho} + 4 \rho^2 U_{k,\rho\rho}
\end{align}
\end{widetext}

\end{document}